\documentclass[pre,final,twocolumn,showkeys,showpacs,groupaddress,preprintnumbers,floatfix]{revtex4-1}

\usepackage{dynlearn}
\usepackage{amsthm}

\newcommand{\msym}{\meassymbol}
\newcommand{\MSym}{\MeasSymbol}

\newcommand{\ProcessLang}{L}
\newcommand{\domain}{\Lambda}
\newcommand{\radius} {R}
\newcommand{\neighborhood} {\eta}

\newcommand{\Localstate} {Local causal state\xspace}
\newcommand{\Localstates} {Local causal states\xspace}
\newcommand{\localstate} {local causal state\xspace}
\newcommand{\localstates} {local causal states\xspace}

\newcommand{\site} {r\xspace}
\newcommand{\point} {(\site, t)\xspace}

\newcommand{\lattice} {\mathcal{L}\xspace}
\newcommand{\stfield} [2] {\ensuremath{\mathbf{\msym}_{#1} ^{#2}}\xspace}
\newcommand{\STField} [2] {\ensuremath{\mathbf{\MSym}_{#1} ^{#2}}\xspace}
\newcommand{\config} {\stfield{t}{} }

\newcommand{\stpoint} {\stfield{t}{\site}}
\newcommand{\STPoint} {\STField{t}{\site}}

\newcommand{\STPointprime} {\STField{t'}{\site'}}

\newcommand{\plc}{\ell^-}
\newcommand{\PLC}{\mathtt{L}^-}
\newcommand{\flc}{\ell^+}
\newcommand{\FLC}{\mathtt{L}^+}
\newcommand{\fhorizon}{h^+}
\newcommand{\phorizon}{h^-}

\newcommand{\causalfield} [2] {\ensuremath{\mathcal{S}_{#1} ^{#2}}\xspace}

\newcommand{\causalpoint} {\causalfield{t}{\site}}

\newcommand{\DPID} {DPID\xspace}

\theoremstyle{definition}
\newtheorem*{defn}{Definition}

\begin{document}

\def\ourTitle{%
Local Causal States \\ and \\
Discrete Coherent Structures
}

\def\ourAbstract{%
Coherent structures form spontaneously in nonlinear spatiotemporal systems and
are found at all spatial scales in natural phenomena from laboratory
hydrodynamic flows and chemical reactions to ocean, atmosphere, and planetary
climate dynamics. Phenomenologically, they appear as key components that
organize the macroscopic behaviors in such systems. Despite a century of
effort, they have eluded rigorous analysis and empirical prediction, with
progress being made only recently. As a step in this, we present a formal
theory of coherent structures in fully-discrete dynamical field theories. It
builds on the notion of structure introduced by computational mechanics,
generalizing it to a local spatiotemporal setting. The analysis' main tool
employs the \localstates, which are used to uncover a system's hidden
spatiotemporal symmetries and which identify coherent structures as
spatially-localized deviations from those symmetries. The approach is
behavior-driven in the sense that it does not rely on directly analyzing
spatiotemporal equations of motion, rather it considers only the spatiotemporal
fields a system generates. As such, it offers an unsupervised approach to
discover and describe coherent structures. We illustrate the approach by
analyzing coherent structures generated by elementary cellular automata,
comparing the results with an earlier, dynamic-invariant-set approach that
decomposes fields into domains, particles, and particle interactions.
}

\def\ourKeywords{%
coherent structures, spatially extended dynamical systems, emergence, symmetry breaking, cellular automata
}

\hypersetup{
  pdfauthor={James P. Crutchfield},
  pdftitle={\ourTitle},
  pdfsubject={\ourAbstract},
  pdfkeywords={\ourKeywords},
  pdfproducer={},
  pdfcreator={}
}

\title{\ourTitle}

\author{Adam Rupe}
\email{atrupe@ucdavis.edu}

\author{James P. Crutchfield}
\email{chaos@ucdavis.edu}

\affiliation{Complexity Sciences Center\\
Physics Department\\
University of California at Davis, One Shields Avenue, Davis, CA 95616}

\date{\today}
\bibliographystyle{unsrt}

\begin{abstract}
\ourAbstract
\end{abstract}



\preprint{\sfiwp{17-09-XXX}}
\preprint{\arxiv{1709.XXXXX}}

\title{\ourTitle}
\date{\today}
\maketitle

\setstretch{1.1}

{\bf
Patterns abound in systems far from equilibrium across all spatial scales, from
planetary and even galactic structures down to the microscopic scales of
snowflakes and bacterial and crystal growth. Most studies of pattern formation,
both theory and experiment, focus on particular classes of human-scale
pattern-forming system and invoke standard bases to describe pattern
organization. This becomes particularly problematic when, for example,
inhomogeneities give rise to relatively more localized patterns, called
coherent structures. Though key to structuring a system's macroscopic behaviors
and causal organization, they have remained elusive for decades. We suggest an
alternative approach that provides constructive answers to the questions of how
to use spacetime fields generated by spatiotemporal systems to extract their
emergent patterns and how to describe them in an objective way.
}

\section{Introduction}
\label{sec:introduction}

Complex patterns are generated by systems in which interactions among their
basic elements are amplified, propagated, and stabilized in a complicated
manner. These emergent patterns present serious difficulties for traditional
mathematical analysis, as one does not know a priori in what representational
basis to describe them, let alone predict them. Notably, analogous difficulties
of describing and predicting the behavior of highly complex systems had been
identified in the early years of computation theory \cite{Mins67} and
linguistics \cite{Chom56}.

A more familiar and perhaps longer-lived example of complex emergent patterns
arises in fluid turbulence \cite{Heis67a}. From its earliest systematic
studies, complex flow patterns were described as linear combinations of
periodic solutions. The maturation of nonlinear dynamical systems theory, though, led
to a radically different view: The mechanism generating complex, unpredictable
behavior was a relatively low-dimensional \emph{strange attractor}
\cite{Lore63,Lore64,Ruel71a}. Using behavior-driven ``state-space
reconstruction'' techniques \cite{Pack80,Take81} this hypothesis was finally
demonstrated \cite{Bran83}. The behavior-driven methods were even extended to
extracting the equations of motion themselves from time series of observations
\cite{Crut87a}. Success in this required knowing an appropriate language with
which to express the equations of motion. Those successes, however,
tantalizingly suggested that behavior-driven methods could let a system's
behavior determine the basis for identifying and describing their emergent
patterns.

To lay the foundations for this and determine what was required for success, a
new approach to discovering patterns generated by complex
systems---\emph{computational mechanics} \cite{Crut88a,Shal98a,Crut12a}---was
developed. It employs mathematical structures analogous to those found in
computation theory to build intrinsic representations of temporal behavior. The
structure of a system's dynamic, the rules of its temporal evolution, are
captured and quantified by the intrinsic representations of computational
mechanics---its \emph{\eMs}. Before this view was introduced, one was tempted
to assume a system's evolution rules were simply its equations of motion. A
hallmark of emergent systems, however, arises exactly when this is not the case
\cite{Crut92c}. While a system's emergent dynamical structure ultimately
derives from the governing equations of motion, arriving at the former from the
latter is typically unfeasible. Similarly, chemistry cannot be considered
simply as ``applied physics'' nor biology, ``applied chemistry'' \cite{Ande72a}.

The use of automata-theoretic constructs lends computational mechanics its
name: it extends statistical mechanics beyond statistics to include
computation-theoretic mechanisms. Operationally, the rise of computer
simulation and numerical analysis as the ``third paradigm'' for physical
sciences provides a research ecosystem that is well-complemented by
computational mechanics, as the latter is a theory built to describe behavior
(data) and, in this, it focuses relatively less on analyzing governing
equations \cite{Crut09c}. The need for behavior-driven
theory---``data-driven'', as some say today---such as computational mechanics
becomes especially apparent in high-dimensional, nonlinear systems.

Patterns abound in systems far from equilibrium across all spatial scales
\cite{Dyke82a,Ball99a,Flak00a}, from galactic structures to planetary---such as
Jupiter's famous Red Spot and similar climatological structures on Earth---down
to the microscopic scales of snowflakes \cite{Naka54a} and bacterial
\cite{Benj95a} and crystal growth \cite{Mark17a}. For imminently practical
reasons, though, most studies of pattern formation, both theory
\cite{Hoyl06a,Cros09a} and experiment, focus on particular classes of
human-scale pattern-forming system, including Rayleigh-B{\'e}nard convection
\cite{Bena01a,Rayl16a,Stei85}, Taylor-Couette flow \cite{Tayl23a,Fens79a}, the
Belousov-Zhabotinsky chemical reaction \cite{Zhab91a,Winf83a}, and Faraday's
crispations \cite{Fara1831a,Kudr98a} to mention several. Often studied under
the rubric of \emph{nonequilibrium phase transitions}
\cite{Reic16a,Atta12a,Hake16a}, these systems are amenable to careful
experimental control and systematic mathematical analysis, facilitated by
imposing idealized boundary conditions.  Nonequilibrium is maintained in these
systems via homogeneous fluxes that give rise to cellular patterns described
and analyzed through global Fourier modes.

While much progress has been made in understanding the instability mechanisms
driving pattern formation and the dynamics of the patterns themselves in
idealized systems \cite{Cros93a,Cros09a,Hoyl06a,Golu03a}, many challenges
remain, especially with wider classes of real world patterns. In particular,
the inescapable inhomogeneities of systems found in nature give rise to
relatively more localized patterns, rather than the cellular patterns captured
by simple Fourier modes. We refer to these localized patterns as \emph{coherent
structures}. There has been intense interest recently in coherent structures in
fluid flows, including structures in geophysical flows \cite{Alls15a, Hall15a},
such as hurricanes \cite{Eman91a, Saps09a}, and in more general turbulent flows
\cite{Holm12a}.

A principled universal description of the organization of such structures does
not exist. So, while we can exploit vast computing resources to simulate models
of ever-increasing mathematical sophistication, analyzing and extracting
insights from such simulations becomes highly nontrivial. Indeed, given the
size and power of modern computers, analyzing their vast simulation outputs can
be as daunting as analyzing any real physical experiment \cite{Crut09c}.
Finally, there is no unique, agreed-upon approach to analyzing and predicting
coherent material structures in fluid flows, for instance \cite{Hadj17a}. Even
today ad hoc thresholding is often used to identify extreme weather events in
climate data, such as cyclones and atmospheric rivers
\cite{Vita97a,Wals97a,Prab15a}. Developing a principled, but general
mathematical description of coherent structures is our focus.

Parallels with contemporary machine learning are worth noting, given the
increasing overlap between these technologies and the needs of the physical
sciences. Imposing Fourier modes as templates for cellular patterns is the
mathematical analog of the technology of (supervised) \emph{pattern
recognition} \cite{Bish06a}. Patterns are given as a finite number of classes
and learning algorithms are trained to assign inputs into these classes by
being fed a large number of labeled training data, which are inputs already
assigned to the correct pattern class.

Computational mechanics, in contrast, makes far fewer structural assumptions
\cite{Crut12a}. As we will see, for discrete spatially extended systems it
makes only modest yet reasonable assumptions about the existence and
conditional stationarity of lightcones in the orbit space of the system. In so
doing, it facilitates identifying representations that are \emph{intrinsic} to
a particular system. This is in contrast with subjectively imposing a
descriptional basis, such as Fourier modes, wavelets, or engineered
pattern-class labels. We say that our subject here is not simply pattern
recognition, but (unsupervised) \emph{pattern discovery}.

To start to address these challenges, we briefly review a particular
spatiotemporal generalization of computational mechanics \cite{Shal03a}. We
adapt it to detect coherent structures in terms of the underlying constituents
from which they emerge, while at the same time providing a principled
description of such structures. The development is organized as follows.
Section~\ref{sec:background} introduces the \localstates, the main tool of
computational mechanics used for coherent structure analysis. We also give an
overview of elementary cellular automata (ECAs), which is the class of
pattern-forming mathematical models we use to demonstrate our coherent
structure analysis.

Section~\ref{sec:CoherentStructures} introduces the computational mechanics of
coherent structures. The dynamical notion of background domains plays a central
role since, after transients die away, the fields produced by spatially
extended dynamical systems can be decomposed into domain regions and coherent
structures embedded in them \cite{Hans90a}. Furthermore, the domains' internal
symmetries typically dictate how the overall spatiotemporal dynamic organizes
itself, including what large-scale patterns may form. More to the point, we
formally define coherent structures with respect to a system's domains.

Crutchfield and Hanson introduced a principled analysis of CA domains and
coherent structures
\cite{Crut88a,Hans90a,Crut91d,Crut92a,Crut93a,Crutchfield&Mitchell94a,Hans95a}.
They defined domains as dynamically invariant sets of spatially statistically
stationary configurations with finite memory. This led to formal methods for
proving that domains were spacetime shift-invariant and so dominant patterns
for a given CA. Having identified these significant patterns, they created
spatial transducers that decomposed a CA spacetime field into domains and
nondomain structures, such as particles and particle interactions
\cite{McTa04a}. We refer to this analysis of CA structures as the
\emph{domain-particle-interaction decomposition} (\DPID). The following extends
\DPID but, for the first time, uses \localstates to define domains and coherent
structures. In this, domains are given by spacetime regions where the
associated \localstates have time and space translation symmetries.

Section~\ref{sec:CAStructures} gives detailed examples for the two main classes
of CA domains---those with explicit symmetries and those with hidden
symmetries. We show empirically that there is a strong correspondence between
domains and structures of elementary CAs identified by \localstates and by the
\DPID approach. For domains, we show that a homogeneous invariant set of
spatial configurations (\DPID domains) produces a \localstate field with a
spacetime symmetry tiling. Since \localstate inference is fully
behavior-driven, it applies to a broader class of spatiotemporal systems than
the \DPID transducers. And so, this correspondence extends both the theory and
application of the coherent structure analysis they engender.

Similar approaches using local causal states have been pursued by others
\cite{Shal06a,Jani07a,Jani10a,Goer12a,Goer13a}. However, as will be elaborated
upon in future work, these underutilize computational mechanics, developing
only a qualitative filtering tool---local statistical complexity---that assists
in subjective visual recognition of coherent structures. Moreover, they provide
no principled way to describe structures and thus cannot, to take one example,
distinguish two distinct types of structures from one another. There have also
been other unsupervised approaches to coherent structure discovery in cellular
automata using information-theoretic measures \cite{Lizi08a,Lizi10a,Flec11a,
Lizi13a}. Recent critiques of employing such measures to determine information
storage and flow and causal dependency \cite{Jame15a,Jame16a} indicate that
these uses of information theory for CAs are still in early development and
have some distance to go to reach the structure-detection performance levels
presented here.

\section{Background}
\label{sec:background}

Modern physics evolved to use group theory to formalize the concept of symmetry
\cite{Tung85a}. The successes in doing so are legion in twentieth-century
fundamental physics. When applied to emergent patterns, though, group-theoretic
descriptions formally describe only their exact symmetries. This is too
restrictive for more general notions---naturally occurring patterns and
structures that are an amalgam of strict symmetry and randomness. Thus, one
appeals to semigroup theory \cite{Laws98a, Kitc86a} to describe partial
symmetries. This use of semigroup algebra is fundamental to automata as
developed in early computation theory \cite{Rhod71a,Holc82a}. In this,
different classes of automata or ``machines'' formalize the concept of
structure \cite{Mins67}. Through the connection with semigroup theory,
structure captured by machines can be seen as a system's generalized symmetries.
The variety of computational model classes \cite{Hopc06a} then becomes an
inspiration for understanding emergent natural patterns \cite{Rhod71a}.

To capture structure in complex physical systems, though, computational
mechanics had to move beyond computation-theoretic automata to probabilistic
representations of behavior. That said, its parallels to semigroups and
automata are outlined in Ref. \cite[Apps. D and H]{Shal98a}, for example. Early
on, the theory was most thoroughly developed in the temporal setting to analyze
\emph{structured} stochastic processes \footnote{Analyzing the statistical
complexity of spatiotemporal dynamics was announced originally, though, in Ref.
\cite[p. 108]{Crut88a}.}. It was also applied to continuous-valued chaotic
systems using the methods \cite{Lind95a} of symbolic dynamics to partition
low-dimensional attractors \cite{Crut88a}. More recently, it has been directly
applied to continuous-time and continuous-value processes
\cite{Marz14a,Marz14c,Marz14e,Crut15a,Marz17b,Riec13a,Riec17a}.

\subsection{Temporal processes, canonical representations}

A \emph{stochastic process} $\Process$ is the distribution of all a system's
allowed behaviors or \emph{realizations} $\ldots \msym_{-2}, \msym_{-1},
\msym_0, \msym_1, \ldots$ as specified by their joint probabilities $\Pr
(\ldots, \MSym_{-2}, \MSym_{-1}, \MSym_0, \MSym_1,...)$. Here, $\MSym_t$ is the
random variable for the outcome of the measurement $\meassymbol_t \in
\MeasAlphabet$ at time $t$, taking values from a finite set $\MeasAlphabet$ of
all possible events. (Uppercase denotes a random variable; lowercase its
value.) We denote a contiguous chain of $\ell$ random variables as
$\MS{0}{\ell} = \MSym_0 \MSym_1 \dotsm \MSym_{\ell -1}$ and their realizations
as $\ms{0}{\ell} = \msym_0 \msym_1 \dotsm \msym_{\ell -1}$. (Left indices are
inclusive; right, exclusive.) We suppress indices that are infinite. We will
often work with \emph{stationary} processes for which $\Pr
\big(\MS{t}{t+\ell}\big) = \Pr \big(\MS{0}{\ell}\big)$ for all $t$ and $\ell$.

The canonical representation for a stochastic process within computational
mechanics is the process' \emph{\eM}. This is a type of stochastic state
machine, commonly known as a \emph{hidden Markov model} (HMM), that consists of
a set $\CausalStateSet$ of \emph{causal states} and transitions between them.
The causal states are constructed for a given process by calculating the
classes determined by the \emph{causal equivalence relation}:
\begin{align*}
\ms{}{t} \; \CausalEquivalence \; \ms{}{t}^\prime \iff
  \Pr ( \MS{t}{} | \MS{}{t} = \ms{}{t} ) =
  \Pr ( \MS{t}{} | \MS{}{t} = \ms{}{t}^\prime )
  ~.
\end{align*}
Operationally, two pasts $\ms{}{t}$ and $\ms{}{t}^\prime$ are causally
equivalent, i.e., belong to the same causal state, if and only if they make the
same prediction for the future. Equivalent states lead to the same future
conditional distribution. Behaviorally, the interpretation is that whenever a
process generates the same future (a conditional distribution), it is
effectively in the same state.

Each causal state $\causalstate \in \CausalStateSet$ is an element of the
coarsest partition of a process' pasts $\{\ms{}{t}: t \in \mathbb{Z}\}$ such
that every $\ms{}{t} \in \causalstate$ has the same predictive distribution:
$\Pr (\MS{t}{} | \ms{}{t}) = \Pr (\MS{0}{} | \cdot)$. The associated random
variable is $\CausalState$. The \emph{$\epsilon$-function} $\epsilon(\ms{}{t})$
maps a past to its causal state: $\epsilon: \ms{}{t} \mapsto \causalstate$. In
this way, it generates the partition defined by the causal equivalence relation
$\CausalEquivalence$. One can show that the causal states are the unique
\emph{minimal sufficient statistic} of the past when predicting the future.
Notably, the causal state set $\CausalStateSet$ can be finite, countable, or
uncountable \cite{Crut92c,Uppe97a,Marz17a}, even if the original process is
stationary, ergodic, and generated by an HMM with a finite set of states.
Reference \cite{Shal98a} gives a detailed exposition and Refs.
\cite{Crut13a,Riec13a,Riec17a} give closed-form calculational tools.

\subsection{Spatiotemporal processes, local causal states}

\newcommand{\state}{\meassymbol}
\renewcommand{\State}{\MeasSymbol}

The \emph{state} $\state$ of a spatiotemporal system specifies the values
$\state^\site$ at \emph{sites} $\site$ of a \emph{lattice} $\lattice$. Assuming
values lie in set $\MeasAlphabet$, a \emph{configuration} $\state \in
\MeasAlphabet^\lattice$ is the collection of values over the lattice sites. If
the values are generated by random variables $\State^\site$, then we have a
\emph{spatial process} $\Pr(\State)$---a stochastic process over the random
variable \emph{field} $\State = \{\State^\site: \site \in \lattice\}$.

A spatiotemporal system, in contrast to a purely temporal one, generates a
process $\Pr(\ldots, \State_{-1}, \State_{0}, \State_{1}, \ldots)$ consisting
of the series of fields $\State_t$. (Subscripts denote time; superscripts
sites.) A realization of a spatiotemporal process is known as \emph{spacetime
field} $\stfield{}{} \in \MeasAlphabet^{\lattice \otimes \mathbb{Z}}$,
consisting of a time series $\state_0, \state_1, \ldots$ of spatial
configurations $\state_t \in \MeasAlphabet^\lattice$. 
$\MeasAlphabet^{\lattice \otimes \mathbb{Z}}$ is the orbit space of the process;
that is, time is added onto the system's state space. The associated spacetime
field random variable is $\STField{}{}$. A \emph{spacetime point} $\stpoint \in
\MeasAlphabet$ is the value of the spacetime field at coordinates
$\point$---that is, at location $\site \in \lattice$ at time $t$. The
associated random variable at that point is $\STField{t}{\site}$.

Being interested in spatiotemporal systems that exhibit spatial translation
symmetries, we narrow consideration to regular spatial lattices with topology
$\lattice = \mathbb{Z}^d$. (As needed, the lattice will be infinite or periodic
along each dimension.)

Purely temporal computational mechanics views the spatiotemporal process
$\Pr(\ldots, \State_{-1}, \State_{0}, \State_{1}, \ldots)$ as a time series
over events with the very large or even infinite alphabet---the configurations
in $\MeasAlphabet^\lattice$. In special cases, one can calculate the temporal
causal equivalence classes and their causal states and transitions from the
time series of spatial configurations, giving the \emph{global \eM}. While
formally well defined, determining the global \eM\ is for all practical
purposes intractable. Some form of simplification is required to make headway.

\subsubsection{Random variable lightcones}

To circumvent this we introduce a different, spatially \emph{local}
representation. This respects and leverages the configurations' spatial nature;
the otherwise unwieldy configuration alphabet $\MeasAlphabet^\lattice$ has
embedded structure. In particular, for systems that evolve under a homogeneous
local dynamic and for which information propagates through the system at a
finite speed, it is quite natural to use lightcones as spatially local notions
of pasts and futures.

\begin{figure}[!htbp]
\centering
\includegraphics[width=\columnwidth]{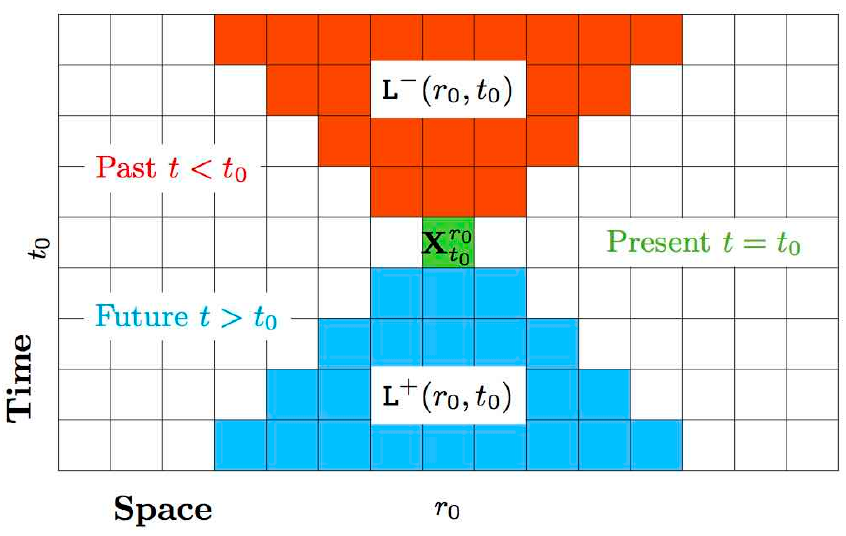}
\caption{Lightcone random variable templates: \textcolor{red}{Past lightcone
	$\PLC (r_0, t_0)$} and \textcolor{cyan}{future lightcone
	$\FLC (r_0, t_0)$} for \textcolor{green!67!black}{present spacetime
	point $\STField{t_0}{r_0}$} in a $1+1$ D field with nearest-neighbor (or
	radius-$1$) interactions.
	}
\label{lightcones}
\end{figure}

Formally, the \emph{past lightcone} $\PLC$ of a spacetime random variable
$\STPoint$ is the set of all random variables at previous times that could
possibly influence it. That is:
\begin{align}
\PLC \point \equiv \big\{ \STPointprime :\; t' \leq t \; \mathrm{and} \;
||\site' - \site|| \leq c(t' - t) \big \}
  ~,
\label{eq:plc}
\end{align}
where $c$ is the finite speed of information propagation in the system.
Similarly, the \emph{future lightcone} $\FLC$ is given as all the random
variables at subsequent times that could possibly be influenced by $\STPoint$:
\begin{align}
\FLC \point \equiv \big \{ \STPointprime :\; t' > t \; \mathrm{and} \;
||\site' - \site|| \leq c ( t - t') \big \}
  ~.
\label{eq:flc}
\end{align}
We include the \emph{present} random variable $\STPoint$ in its past lightcone,
but not its future lightcone. An illustration for one-space and time ($1 + 1
$D) fields on a lattice with nearest-neighbor (or \emph{radius}-$1$)
interactions is shown in Fig~\ref{lightcones}. We use $\PLC$ to denote the
random variable for past lightcones with realizations $\plc$; similarly, $\FLC$
those with realizations $\flc$ for future lightcones.

The choice of lightcone representations for both local pasts and futures is
ultimately a weak-causality argument: influence and information propagate
locally through a spacetime site from its past lightcone to its future
lightcone. A sequel \cite{Rupe17c} goes into more depth, exploring this choice
and possible variations. For now, we work with the given assumptions.

Using lightcones as local pasts and futures, generalizing the causal
equivalence relation to spacetime is now straightforward. Two past lightcones
are causally equivalent if they have the same distribution over future
lightcones:
\begin{align}
\plc_i \; \CausalEquivalence \; \plc_j \iff \Pr \big(\FLC | \plc_i \big) = \Pr \big(\FLC | \plc_j \big)
\label{eq:LocalCEquiv}
  ~.
\end{align}
This \emph{local causal equivalence relation} over lightcones implements an
intuitive notion of \emph{optimal local prediction} \cite{Shal03a}. At some
point $\stpoint$ in spacetime, given knowledge of all past spacetime points
that could possibly affect $\stpoint$---i.e., its past lightcone
$\plc \point$---what might happen at all subsequent spacetime points that
could be affected by $\stpoint$---i.e., its future lightcone
$\flc \point$?

The equivalence relation induces a set $\CausalStateSet$ of \emph{local causal
states} $\causalstate$. A functional version of the equivalence relation is
helpful, as in the pure temporal setting, as it directly maps a given past
lightcone $\ell^-$ to the equivalence class $[\ell^-]$ of which it is a member:
\begin{align*}
\epsilon(\ell^-) & = [\ell^-] \\
  & = \{\ell^{-'}: \ell^- \sim_\epsilon \ell^{-'} \}
\end{align*}
or, even more directly, to the associated local causal state:
\begin{align*}
\epsilon(\ell^-) = \causalstate_{\ell^-}
  ~.
\end{align*}

Closely tracking the standard development of temporal computational mechanics
\cite{Shal98a}, a set of results for spatiotemporal processes parallels those
of temporal causal states \cite{Shal03a}. For example, one concludes that
\localstates are minimal sufficient statistics for optimal local prediction.
Moreover, the particular local prediction uses lightcone-shaped random-variable
templates, associated with local causality in the system. Specifically, the
future follows the past and information propagates at a finite speed. Thus,
\localstates do not detect direct causal relationships---say, as reflected in
learning equations of motion from data. Rather, they exploit an intrinsic
causality in the system in order to discover emergent spacetime structures.

As an aside, if viewed as a form of data-driven machine learning, our
coherent-structure theory, implemented using either \DPID or local causal
states, allows for unsupervised image-segmentation labeling of spatiotemporal
structures. We should emphasize that this is \emph{spacetime segmentation} and
not a general image segmentation algorithm \cite{Bish06a}, since it works \emph{only} in
systems for which local causality exists and for which lightcone templates are
well defined.

\subsubsection{Causal state filtering}

As in purely-temporal computational mechanics, the local causal equivalence
relation Eq. (\ref{eq:LocalCEquiv}) induces a partition over the space of
(infinite) past lightcones, with the \localstates being the equivalence
classes. We will use the same notation for local causal states as was used for
temporal causal states above, as there will be no overlap later:
$\CausalStateSet$ is the set of local causal states defined by the local causal
equivalence partition, $\CausalState$ denotes the random variable for a local
causal state, and $\causalstate$ for a specific realized causal state. The local
$\epsilon$-function $\epsilon(\plc)$ maps past lightcones to their \localstates
$\epsilon :\plc \mapsto \causalstate$, based on their conditional distribution
over future lightcones.

For spatiotemporal systems, a first step to discover emergent patterns applies
the local $\epsilon$-function to an entire spacetime field to produce an
associated \emph{\localstate field} $\causalfield{}{} =
\epsilon(\stfield{}{})$. Each point in the \localstate field is a \localstate
$\causalpoint = \causalstate \in \CausalStateSet$.

The central strategy here is to extract a spatiotemporal process' pattern and
structure from the \localstate field. The transformation $\causalfield{}{} =
\epsilon(\stfield{}{})$ of a particular spacetime field realization
$\stfield{}{}$ is known as \emph{causal state filtering} and is implemented as
follows. For every spacetime coordinate $\point$:
\begin{enumerate}
\setlength{\topsep}{-3pt}
\setlength{\itemsep}{-3pt}
\setlength{\parsep}{-3pt}
\setlength{\labelwidth}{5pt}
\setlength{\itemindent}{0pt}
\item At $\stpoint$ determine its past lightcone $\PLC \point = \plc$;
\item Form its local predictive distribution $\Pr(\FLC | \plc)$;
\item Determine the unique \localstate $\causalstate \in \CausalStateSet$
	to which it leads; and
\item Label the \localstate field at point $\point$ with $\causalstate$:
	$\causalpoint = \causalstate$.
\end{enumerate}
Notice the values assigned to $\causalfield{}{}$ in step 4 are simply the
labels for the corresponding \localstates. Thus, the \localstate field is a
\emph{semantic field}, as its values are not measures of any quantity, but
rather labels for equivalence classes of local dynamical behaviors as in the
\emph{measurement semantics} introduced in Ref. \cite{Crut91b}.

In practice, there are inference details involved in causal filtering which we
discuss more in Ref. \cite{Rupe17c}. The main inference parameters are the
finite lightcone horizons $\phorizon$ and $\fhorizon$, as well as the speed of
information propagation $c$. For cellular automata $c$ is simply the radius
$\radius$ of local neighborhoods; see below. These parameters determine the
shape of the lightcone templates that are extracted from spacetime fields.

Causal state filtering will be used shortly in Sec.~\ref{sec:CoherentStructures} to analyze spacetime domains and coherent structures. For each case we will give the past and future lightcone horizons used. But first we must introduce prototype spatial dynamical systems to study.

\subsection{Cellular automata}

The spatiotemporal processes whose structure we will analyze are
deterministically generated by cellular automata. A \emph{cellular automaton}
(CA) is a a fully-discrete spatially-extended dynamical system with a regular
spatial lattice in $d$ dimensions $\lattice = \mathbb{Z}^d$, consisting of
local variables taking values from a discrete alphabet $\MeasAlphabet$ and
evolving in discrete time steps according to a \emph{local dynamic} $\phi$.
Time evolution of the value at a site on a CA's lattice depends only on values
at sites within a given \emph{radius} $\radius$. The collection of all sites
within radius $\radius$ of a point $\stpoint$, including $\stpoint$ itself, is
known as the point's \emph{neighborhood} $\neighborhood(\stpoint)$:
\begin{align*}
\neighborhood(\stpoint)
  = \{\stfield{t}{\site'} \; : \; ||\site - \site'|| \leq R,
  \site, \site' \in \lattice\}
  ~.
\end{align*}
The neighborhood specification depends on the form of the lattice distance
metric chosen. The two most common neighborhoods for regular lattice
configurations are the Moore and von Neumann neighborhoods, defined by the
Chebyshev and Manhattan distances in $\lattice$, respectively.

The \emph{local} evolution of a spacetime point is given by:
\begin{align*}
\stfield{t+1}{\site} = \phi \big(\neighborhood(\stpoint)  \big) ~,
\end{align*}
and the \emph{global} evolution $\Phi : \MeasAlphabet^\lattice \rightarrow
\MeasAlphabet^\lattice$ of the spatial field is given by:
\begin{align}
\stfield{t+1}{} = \Phi(\config)
  ~.
\label{eq:GlobalCAEv}
\end{align}
For example, this might apply $\phi$ in parallel, simultaneously to all
neighborhoods on the lattice. Although, other local update schemes are
encountered.

As noted, CAs are fully discrete dynamical systems. They evolve an initial
spatial configuration $\state_0 \in \MeasAlphabet^\lattice$ according to Eq.
(\ref{eq:GlobalCAEv})'s dynamic. This generates an \emph{orbit}
$\stfield{0:t}{} = \{\stfield{0}{}, \stfield{1}{}, \ldots \stfield{t-1}{}\} \in 
\MeasAlphabet^{\lattice \otimes \mathbb{Z}}$.
Usefully, dynamical systems theory classifies a number of orbit types. Most
basically, a \emph{periodic orbit} repeats in time:
\begin{align}
\stfield{t+p}{} & = \Phi^p(\stfield{t}{}) \nonumber \\
   & = \stfield{t}{} ~,
\label{eq:PeriodicOrbit}
\end{align}
where $p$ is its \emph{period}---the smallest integer for which this holds.
A \emph{fixed point} has $p = 1$ and a \emph{limit cycle} has finite $p > 1$.
An \emph{aperiodic orbit} has no finite $p$; a behavior that can occur only on
infinite lattices.

Since CA states are spatial configurations an orbit $\stfield{0:t}{}$ is a spacetime field. These orbits constitute the spatiotemporal processes of
interest in the following.

\subsection{Elementary CAs}

The prototype spatial systems we use to demonstrate coherent structure analysis
are the \emph{elementary cellular automata} (ECAs) that have a one-dimensional
spatial lattice $\lattice = \mathbb{Z}$ and local random variables taking
binary values $\MeasAlphabet = \{0,1\}$. Thus, ECA spatial configurations
$\config \in \MeasAlphabet^\mathbb{Z}$ are strings of 0s and 1s. Equation
(\ref{eq:GlobalCAEv})'s time evolution is implemented by simultaneously
applying the local dynamic (or \emph{lookup table}) $\phi$ over radius-$1$
neighborhoods $\neighborhood (\stfield{t}{r}) = \stfield{t}{r-1} \stfield{t}{r}
\stfield{t}{r+1}$:
\begin{align*}
\begin{tabular}{c c c | c}
\multicolumn{3}{c|}{$\eta$} & $O_\eta = \phi(\eta)$ \\
\hline
1 & 1 & 1 & $O_7$\\
1 & 1 & 0 & $O_6$ \\
1 & 0 & 1 & $O_5$ \\
1 & 0 & 0 & $O_4$ \\
0 & 1 & 1 & $O_3$ \\
0 & 1 & 0 & $O_2$ \\
0 & 0 & 1 & $O_1$ \\
0 & 0 & 0 & $O_0$
\end{tabular}
  ~,
\end{align*}
where each output $O_\eta = \phi(\eta) \in \MeasAlphabet$ and the $\eta$s are
listed in lexicographical order. There are $2^8 = 256$ possible lookup tables,
as specified by the string of output bits: $O_7 O_6 O_5 O_4 O_3 O_2 O_1 O_0$. A
specific ECA lookup table is often referred to as an ECA \emph{rule} with a
\emph{rule number} given as the binary integer $o_7 o_6 o_5 o_4 o_3 o_2 o_1 o_0
\in [0,255]$. For example, ECA 172's lookup table has output bit string
$10101100$. Arguably, ECAs are the simplest pattern-forming spatially extended
dynamical system \cite{Wolf83}.

Over the years, CAs have been designed as distributed implementations of
various kinds of computation. In this, one studies specific combinations of
initial conditions and CA rules. For example, over a restricted set of
initial configurations ECA $110$ is computation universal, a capability it
embodies via its coherent structures \cite{Cook04a}. Here, though, we are
interested in typical spatiotemporal behaviors generated by ECAs. Practically
speaking, this means analyzing spacetime fields that are generated from random
initial conditions under a given ECA rule. In short, our studies will randomly
sample the space of field configurations generated by given ECA rules. It is
convenient to consider boundary conditions consistent with spatial translation
symmetry. For numerical simulations, as we used here, this means using periodic
boundary conditions.

To close, we note the relationship between past lightcones and a CA's local
dynamic $\phi$. The $i^{th}$-order lookup table $\phi^i$ maps the radius $\radius =
i \cdot c$ neighborhood of a site to that site's value $i$ time steps in the
future. Said another way, a spacetime point $\stfield{t+i}{\site}$ is
completely determined by the radius $\radius = i\cdot c$ neighborhood $i$
time-steps in the past according to $\phi^i\big (\eta^{i\cdot c}
(\stpoint)\big)$. To fill out the elements of $\phi^i$,
apply $\phi$ to all points of $\eta^{i\cdot c}$ to produce $\eta^{(i-1)\cdot
c}$ and so on until $\eta^{0} = \stpoint$ is reached. This is what we call
the \emph{lookup table cascade}, the elements of which are finite-depth past
lightcones.

\subsection{Automata-theoretic CA evolution} 

For cellular automata in one spatial dimension, such as ECAs, configurations
$\config \in \MeasAlphabet^{\mathbb{Z}}$ are strings over the alphabet
$\MeasAlphabet$. Rather than study how a CA evolves individual configurations,
it is particularly informative to investigate how CAs evolve sets. This is a
key step \DPID uses in discovering a CA's emergent patterns \cite{Hans90a}.
We are particularly interested in how the spatial structure in a CA's
configurations evolve. To monitor this, we use automata-theoretic
representations of sets of spatial configurations.

Sets of strings recognized by finite-state machines are called \emph{regular
languages}. Any regular language $\ProcessLang$ has a unique minimal
finite-state machine $M(\ProcessLang)$ that recognizes or generates it
\cite{Hopc06a}. These automata are particularly useful since they give a finite
mathematical representation of a typically infinite set of configurations that
are regular languages.

To explore how a CA evolves languages we establish a dynamic that evolves
machines. This is accomplished via finite-state transducers. Transducers are a
particular type of input-output machine that maps strings to strings
\cite{Broo89a}. This is exactly what the global dynamic of a CA does
\cite{Wolf84a}. As a mapping from a configuration $\stfield{t}{}$ at time $t$
to one $\stfield{t+1}{}$ at time $t+1$, it is also a map on a configuration set
$\ProcessLang_t$ from one time to the next $\ProcessLang_{t+1}$:
\begin{align}
\ProcessLang_{t+1} = \Phi(\ProcessLang_t)
	~.
\label{eq:CAMapsSets}
\end{align}

A CA's global dynamic $\Phi$, though, can be represented as a finite-state
transducer $\mathrm{T}_\Phi$ that evolves a set of configurations represented
by a finite-state machine. This is the \emph{finite machine evolution} (FME)
operator \cite{Hans90a}. Its operation composes the CA transducer
$\mathrm{T}_\Phi$ and finite-state machine $M(\ProcessLang_t)$ to get the
machine $M_{t+1} = M(\ProcessLang_{t+1})$ describing the set
$\ProcessLang_{t+1}$ of spatial configurations at the next time step:
\begin{align}
M_{t+1} = \min \big( \mathrm{T}_\Phi \circ M(\ProcessLang_t) \big)
  ~.
\label{eq:FME}
\end{align}
Here, $\min (M)$ is the automata-theoretic procedure that minimizes the number
of states in machine $M$. While not entirely necessary for language evolution,
the minimization step is helpful when monitoring the complexity of
$\ProcessLang_t$. The net result is that Eq. (\ref{eq:FME}) is the
automata-theoretic version of Eq. (\ref{eq:CAMapsSets})'s set evolution
dynamic. Analyzing how the FME operator evolves configuration sets of different
kinds is a key tool in understanding CA emergent patterns.

\section{Domains and Coherent Structures}
\label{sec:CoherentStructures}

The following develops our theory of coherent structures and then demonstrates
it by identifying patterns in ECA-generated spacetime fields. The theory builds
off the conceptual foundation laid out by \DPID in which structures, such
as particles and their interactions, are seen as deviations from spacetime
shift-invariant domains. The new \localstate formulation differs from \DPID
in how domains and their deviations are formally defined and identified. The
two distinct approaches to the same conceptual objective complement and
inform one another, lending distinct insight into the patterns and regularity
captured by the other.

We begin with an overview of \DPID CA pattern analysis and then present the new
formulation of domains based on \localstates. Generalizing \DPID particles,
coherent structures are then formally defined as particular deviations from
domains. Specifically, coherent structures are defined through \emph{semantic
filters} that use either the \localstate field $\causalfield{}{} =
\epsilon(\stfield{}{})$ or the \DPID domain-transducer filter described
shortly. CA coherent structures defined via the latter are \DPID particles.
Defining particles using \localstates, in contrast, extends
domain-particle-interaction analysis to a broader class of spatiotemporal
systems for which \DPID transducers do not exist. Due to this improvement, in
the \localstates analysis we adopt the terminology of ``coherent structures'' over ``particles''.

\subsection{Domains}

The approach to coherent structures begins with what they are not. Generally,
structures are seen as deviations from spatially and temporally statistically
homogeneous regions of spacetime. These homogeneous regions are generally
called \emph{domains}, alluding to solid state physics. They are the background
organizations above which coherent structures are defined.

\subsubsection{Structure from breaking symmetries}

Structure is often described as arising from \emph{broken symmetries}
\cite{Ande72a,Nico77a,Ball99a,Hake83a,Seth92a,Cros93a,Hoyl06a,Livi12a}. Though
key to our development, broken symmetry is a more broadly unifying mechanism in
physics. Care, therefore, is required to precisely distinguish the nature of
broken symmetries we are interested in. Specifically, our formalism seeks to
capture coherent structures as \emph{temporally-persistent, spatially-localized
broken symmetries}.

Drawing contrasts will help delineate this notion of coherent structure from
others associated with broken symmetries. Equilibrium phase transitions also
arise via broken symmetries. There, the degree of breaking is quantified by an
\emph{order parameter} that vanishes in the symmetric state. A transition
occurs when the symmetry is broken and the order parameter is no longer zero
\cite{Seth92a}.

This, however, does not imply the existence of coherent structures. When the
order parameter is global and not a function of space, symmetry is broken
globally, not locally. And so, the resulting state may still possess additional
global symmetries. For example, when liquids freeze into crystalline solids,
continuous translational symmetry is replaced by a discrete translational
symmetry of the crystal lattice---a global symmetry.

Similarly, the primary bifurcation exhibited in nonequilibrium phase
transitions occurs when the translational invariance of an initial homogeneous
field breaks \cite{Cros93a,Hoyl06a}. It is often the case, though, as in
equilibrium, that this is a continuous-to-discrete symmetry breaking, since the
cellular patterns that emerge have a discrete lattice symmetry. To be concrete,
this occurs in the conduction-convection transition in Rayleigh-B{\'e}nard flow.
The convection state just above the critical Rayleigh number consists of
convection cells patterned in a lattice \cite{Bena01a,Chan68a}. In the language
used here, the above patterns arise as a change of domain structure, not the
formation of coherent structures. Coherent structures, such as topological defects \cite{Cros93a,Sigg81a}, form at higher Rayleigh numbers when the discrete cellular symmetries are locally broken.

Describing domains, their use as a baseline for coherent structures, and how
their own structural alterations arise from global symmetry breaking
transitions delineates what our coherent structures are not. To make positive
headway, we move on to a direct formulation, starting with how they first
appeared in the original \DPID and then turning to express them via
\localstates.

\subsubsection{\DPID patterns}

Domains of one-dimensional cellular automata were defined in \DPID pattern
analysis \cite{Hans90a,Crut91d,Crut92a,Crut93a,Hans95a} as configuration sets
that, when evolved under the system's dynamic, produce spacetime fields that
are time- and space-shift invariant. Formally, the computational mechanics of
spacetime fields was augmented with concepts from dynamical systems---invariant
sets, basins, attractors, and the like---adapted to describe organization in
the CA infinite-dimensional state space $\MeasAlphabet^\infty$. There, a
\emph{domain} $\domain \subseteq \MeasAlphabet^\infty$ of a CA $\Phi$ is a set
of spatial configurations with:
\begin{enumerate}
\item \emph{Temporal invariance}: $\domain$ is mapped onto itself by the
	global dynamic $\Phi$:
\begin{align}
\Phi^{\widehat{p}} (\domain) = \domain
  ~,
\label{eq:DomainTemporalInvar}
\end{align}
for some finite time $\widehat{p}$; and
\item \emph{Spatial invariance}: $\domain$ is mapped onto itself by the
	spatial-shift dynamic $\sigma$:
\begin{align}
\sigma^s (\domain) = \domain
  ~,
\label{eq:DomainSpatialInvar}
\end{align}
for some finite distance $s$.
\end{enumerate}
The smallest $\widehat{p}$ for which the temporal invariance of Eq.
(\ref{eq:DomainTemporalInvar}) holds gives the domain's \emph{recurrence
time}.  Similarly, the smallest $s$ is domain's \emph{spatial period}. In this
way, a domain $\Lambda$ consists of $\widehat{p}$ temporal phases, each its own
spatial language: $\Lambda = \{\Lambda_1, \Lambda_2, \ldots,
\Lambda_{\widehat{p}} \}$. In the terminology of symbolic dynamics
\cite{Lind95a}, each temporal phase $\Lambda_i$ is a shift space
$\mathcal{X}_{\Lambda_i} \subseteq \MeasAlphabet^{\mathbb{Z}}$ (spatial shift
invariance) such that the CA dynamic $\Phi^{\widehat{p}}$ is a conjugacy from
$\mathcal{X}_{\Lambda_i}$ to itself (temporal invariance). 

An ambiguity arises here between $\Lambda$'s recurrence time $\widehat{p}$ and its
\emph{temporal period} $p$. For a certain class of CA domain (those with explicit symmetries, see 
Sec~\ref{sec:CAStructures} A), 
the domain states $\state \in \domain$ are periodic orbits of the CA, with orbit period 
equal to the domain period, $\state = 
\Phi^p(\state)$. More generally, the recurrence time $\widehat{p}$ is the time required 
for the domain to return to the spatial language temporal phase it started in. That is,
if initially in phase $\domain_i$, $\widehat{p}$ is the number of time steps required to return 
to $\domain_i$. The temporal period of the domain, in contrast, is the number of time 
steps required not just to return to $\domain_i$, but to return to $\domain_i$ in the same 
spatial phase it started in. Thus $\widehat{p} \leq p$. Determining $p$ involves 
examining how $\phi$ interacts with $\domain$, rather than $\Phi$.

Once a domain $\domain$ is found it is straightforward to use $\phi$ to
construct a \DPID \emph{spacetime machine} that describes $\Lambda$'s allowed
spacetime regions \cite{Hans95a}. We refer to a CA domain that is a regular
language as a \emph{regular domain}. Roughly speaking, this captures the notion
of a spatial (or a spacetime) region generated by a locally finite-memory
process.

How does one find domains for a given CA in the first place? While there are no
general analytic solutions to Eq. (\ref{eq:DomainTemporalInvar}), checking that
a candidate language $\ProcessLang$ is invariant under the dynamic is
computationally straightforward using Eq. (\ref{eq:FME})'s FME operator, if
potentially compute intensive. The FME operator is repeated $\widehat{p}$ times
to construct $M(\Phi^{\widehat{p}} \ProcessLang)$ to symbolically---that is,
exactly---check whether a candidate language is periodic under the CA dynamic:
$M(\ProcessLang) \simeq M(\Phi^{\widehat{p}} \ProcessLang)$, where we compare
up to isomorphism implemented using automata minimization. Spatial translation
invariance then requires checking that $M(\ProcessLang)$ has a single strongly
connected set of recurrent states. This is a subtle point, as a corollary to
the Curtis-Hedlund-Lyndon theorem \cite{Hedl69a} states that every image of
cellular automata is a shift space and thus described by a strongly connected
automata \cite{Cecc10a}. This however, concerns evolving single configurations,
whereas the FME operator evolves configuration sets. Thus, a single strongly
connected set of recurrent states as output of FME is nontrivial and shows that
set consists of spatially homogeneous configurations.

Using FME, one can ``guess and check'' candidate domains. This can be automated
since candidate regular domain machines can be exactly enumerated in increasing
number of states and transitions \cite{John10a}. Fortunately, too, not all
possible candidates need be considered. Loosely speaking, one may think of
domain languages as ``spatial \eMs''. Equation (\ref{eq:DomainSpatialInvar})'s
domain spatial-shift invariance establishes \eM properties (e.g. minimality and
unifilarity) for candidate languages $\ProcessLang$. This substantially
constrains the space of possible languages, as well as introduces the
possibility of using \eM inference algorithms \cite{Stre13a} when working with
empirical spacetime datasets.  Additional constraints can further reduce search
time, but these details need not concern us here. 

Once a CA's domains $\domain^0, \domain^1, \ldots$ are discovered, they can be
used to create a \emph{domain transducer} $\tau$ that
identifies which of configuration $\state$'s sites are in which domain
and which are not in any domain \cite{McTa04a}. For a given 1+1 dimension spacetime field 
$\stfield{}{}$, each of its spatial configurations $\state = \stfield{t}{}$ are 
scanned by the transducer, with output $T_t = \tau(\state)$. Although the transducer maps 
strings to strings, the full spacetime field can be filtered with $\tau$ by collecting
the outputs of each configuration in time order to produce the \emph{domain transducer filter 
field} of $\stfield{}{}$: $T = \tau(\stfield{}{}$). 

Sites $\stfield{t}{r}$
``participating'' in domain $\domain^i$ are labeled $i$ in the transducer
field. That is:
\begin{align*}
T_t ^r = \tau(\stfield{t}{})^r = i
  ~.
\end{align*}
Other sites are similarly labeled by the particular way in which they deviate
from domains. One or several sites, for example, can indicate transitions from
one domain temporal phase or domain type to another. If that happens in a way
that is localized across space, one refers to those sites as participating in a
CA \emph{particle}. \emph{Particle interactions} can also be similarly
identified. Reference \cite{Hans90a} gives describes how this is carried out.

In general, a stack automaton is needed to perform this domain-filtering task,
but it may be efficiently approximated using a finite-state transducer
\cite{McTa04a}. 

This filter allows us not only to formally define CA domains,
the transducer allows for site-by-site identification of domain regions and
thus also sites participating in nondomain patterns. In this way and in a
principled manner, one finds localized deviations from domains---these are our
candidate coherent structures.

Originally, this was called \emph{cellular automata computational mechanics}.
Since then, other approaches to spatiotemporal computational mechanics
developed, such as \localstates. We now refer to the above as \DPID pattern
analysis.

\subsubsection{\Localstate patterns}

\DPID pattern analysis formulates domains directly in terms of how a system's
dynamic evolves spatial configurations. That is, domains are sets of
structurally homogeneous spatial configurations that are invariant under
$\Phi$. While this is appealing in many ways, it can become cumbersome in more
complex spatiotemporal systems.

Let's be clear where such complications arise. On the one hand, estimating a
CA's rule $\phi$ and so building up $\Phi$ is straightforwardly implemented by
scanning a spacetime field for neighborhoods and next-site values. This sets up
\DPID with what it needs. On the other, there are circumstances in which a
finite-range rule $\phi$ is not available, leaving \DPID mute. This can occur
even in very simple settings. The simplest with which we are familiar arises in
hidden cellular automata---the \emph{cellular transducers} of Ref.
\cite{Crut91e}. There, perhaps somewhat surprisingly, ECA evolution observed
through other radius-$1$ rule tables generate spacetime data that \emph{no
finite-radius} CA can generate.

For these reasons and to develop methods for even more complicated
spatiotemporal systems where the FME operator cannot be applied, we now develop
a companion approach. Just as the causal states help discover structure from a
temporal process, we would like to use the \localstates to discover structure,
in the more concrete sense of coherent structures, directly from spacetime
fields. To do so, we start with a precise formulation of domains in terms of
\localstates. Since \localstates apply in arbitrary spatial dimensions,
the following addresses general $d$-dimensional cellular automata. In this,
index $n \in \{1,2,\ldots,d\}$ identifies a particular spatial coordinate.

A simple but useful lesson from \DPID is that domains are special (invariant)
subsets of CA configurations. Since they are deterministically generated, a
CA's spacetime field is entirely specified by the rule $\phi$, the initial
condition $\stfield{0}{}$, and the boundary conditions. Here, in analyzing a
CA's behavior, $\phi$ is fixed and we only consider periodic boundary
conditions. This means for a given CA rule, the spacetime field is entirely
determined by $\stfield{0}{}$. If it belongs to a domain---$\stfield{0}{} \in
\domain^i$---all subsequent configurations of the spacetime field will, by
definition, also be in the domain---$\stfield{t}{} = \Phi^t(\stfield{0}{})  \in
\domain^i$. In this sense a domain $\domain \subseteq \MeasAlphabet^{\lattice}$
is a subset of a CA's allowed behaviors: $\domain \subseteq
\Phi^t(\MeasAlphabet^{\lattice})$, $t = 1, 2, 3, \ldots$.

Lacking prior knowledge, if one wants to use \localstates to discover a CA's
patterns, their reconstruction should be performed on \emph{all} of a CA's
spacetime behavior $\Phi^t(\MeasAlphabet^{\lattice})$. This gives a complete
sampling of spacetime field realizations and so adequate statistics for good
\localstate inference. Doing so leaves one with the full set of \localstates
associated with a CA. Since domains are a subset of a CA's behavior, they must
be described by some special subset of the associated \localstates. What are
the defining properties of this subset of states which define them as one or
another domain?

The answer is quite natural. The defining properties of \localstates associated
with domains are expressed in terms of symmetries. For one-dimensional CAs
these are time and space translation symmetries. In general, alternative
symmetries may be considered as well, such as rotations, as appropriate to
other settings. Such symmetries are directly probed through causal filtering.

Consider a domain $\domain$, the local causal states $\CausalStateSet$ induced
by the local causal equivalence relation $\sim_\epsilon$ over spatiotemporal
process $\STField{}{}$, and the \localstate field $\causalfield{}{} =
\epsilon(\stfield{}{})$ over realization $\stfield{}{}$. Let $\sigma_p$ denote 
the \emph{temporal shift operator} that shifts a spacetime field 
$\stfield{}{}$ $p$ steps along the time dimension. This translates a 
point $\stpoint$ in the spacetime field as: $\sigma_p (\stfield{}{})_t^r = 
\stfield{t+p}{r}$. Similarly, let $\sigma^{s_n}$ denote the \emph{spatial 
shift operator} that shifts a spacetime field $\stfield{}{}$ by $s_n$ steps 
along the $n^\text{th}$ spatial dimension. This translates a spacetime point $\stpoint$ 
as: $\sigma^{s_n}(\stfield{}{})_t^r = \stfield{t}{r'}$, where $r'_n = r_n + s_n$.

\begin{defn}
A \emph{pure domain field} $\stfield{\domain}{}$ is a realization such that
$\sigma_{p}$ and the set of spatial shifts $\{\sigma^{s_n}\}$ applied
to $\causalfield{\domain}{} = \epsilon(\stfield{\domain}{})$ form a symmetry
group. The generators of the symmetry group consist of the following translations: 
\begin{enumerate}
\item
\emph{Temporal invariance}: For some finite time shift $p$ the domain causal state field is invariant:
\begin{align}
\sigma_p (\causalfield{\domain}{}) = \causalfield{\domain}{}
  ~,
\label{eq:LCSTempInvar}
\end{align}
and:
\item
\emph{Spatial invariance}: For some finite spatial shift $s_n$ in each
spatial coordinate $n$ the domain causal state field is invariant:
\begin{align}
\sigma^{s_n} (\causalfield{\domain}{}) = \causalfield{\domain}{}
  ~.
\label{eq:LCSSpatialInvar}
\end{align}
\end{enumerate}
The symmetry group is completed by including these translations' inverses,
compositions, and the identity null-shift $\sigma_\mathbf{0}(\stfield{}{})_t^r = \stpoint$.
The set $\CausalStateSet_\domain \subseteq\CausalStateSet$ is $\domain$'s 
\emph{domain \localstates}: $\CausalStateSet_\domain = \{
\big(\causalfield{\domain}{}\big)_t^{\site} : ~ t \in \mathbb{Z},
\site \in \lattice\}$.
\end{defn}

The smallest integer $p$ for which the temporal invariance of Eq.
(\ref{eq:LCSTempInvar}) is satisfied is $\domain$'s \emph{temporal period}. The
smallest $s_n$ for which Eq.  (\ref{eq:LCSSpatialInvar})'s spatial invariance
holds is $\domain$'s \emph{spatial period} along the $n^\text{th}$ spatial
coordinate. 

The domain's \emph{recurrence time} $\widehat{p}$ is the smallest time shift
that brings $\causalfield{\domain}{}$ back to itself when also combined with
finite spatial shifts. That is, $\sigma^j
\sigma_{\widehat{p}}(\causalfield{\domain}{}) = \causalfield{\domain}{}$ for
some finite space shift $\sigma^j$. If $\widehat{p} > 1$, this implies there
are distinct tilings of the spatial lattice at intervening times between
recurrence. The distinct tilings then correspond to $\domain$'s temporal
phases: $\domain = \{\domain_1, \domain_2, \dots, \domain_{\widehat{p}}\}$. For
systems with a single spatial dimension, like the ECAs, the spatial symmetry
tilings are simply $(\causalfield{\domain}{})_t = \cdots w \cdot w \cdot w
\cdots = w^\infty$, where $w = (\causalfield{\domain}{})_t^{i: i+s}$. Each
domain phase $\domain_i$ corresponds to a unique tiling $w_i$. 

For both the \DPID and \localstate formulations of domain we use the notation
$p$ for temporal period, $s$ for spatial period, and $\widehat{p}$ for
recurrence time. While there is as yet no theoretical justification or a priori
reason to assume these are the same, we anticipate the empirical correspondence
between the two distinct formulations of domain when applied to CAs, as seen
below in Sec~\ref{sec:CAStructures}. This also relieves us and the reader of
excess notation. 

Consider a contiguous region ${\mathcal{R}}_\domain\subset \lattice \otimes \mathbb{Z}$ in 
$\causalfield{}{} = \epsilon(\stfield{}{})$ for
spacetime field $\stfield{}{}$ for which all points $\causalpoint$ in the region are domain
\localstates: $\causalpoint \in \CausalStateSet_{\domain} \; , \; (r,t) \in \mathcal{R}_\domain$. The space and time shift
operators over the region obey the symmetry groups of pure domain fields.  Such
regions, over both $\stfield{}{}$ and $\causalfield{}{} = \epsilon(\stfield{}{})$, are \emph{domain regions}.

Once a CA's \localstates are identified, one can track unit-steps in space and
in time over \localstate fields $\causalfield{}{}$ to construct a
\emph{spacetime machine} (an automaton) consisting of the \localstates and
their allowed transitions. That is, if $\causalpoint = \causalstate$ and
$\causalfield{t}{r+1} = \causalstate^\prime$, then if one moves from $(r,t)$ to
$(r+1,t)$ in the spacetime field, one sees a spatial transition between
$\causalstate$ and $\causalstate^\prime$ in the spacetime machine. Similarly, a
temporal transition between $\causalstate$ and $\causalstate^\prime$ is seen if
$\causalpoint = \causalstate$ and $\causalfield{t+1}{r} = \causalstate^\prime$.

The symmetry tiling of domain states determines a particular substructure in
the full spacetime machine. Specifically, for each state $\causalstate \in
\CausalStateSet_{\domain}$ there is a transition leading to state
$\causalstate^\prime \in \CausalStateSet_{\domain}$ if
$(\causalfield{\domain}{})_t^{\site} = \causalstate$ and
$\sigma(\causalfield{\domain}{})_t^{\site} = \causalstate^\prime$, where
$\sigma$ generically denotes a unit shift in time or space. This \emph{domain
submachine} is the analog of the \DPID domain spacetime machine \cite{Hans95a}.
In fact, in all known cases the two spacetime domain machines are identical, up
to isomorphism.

With this set-up, discovering the domains of a spatiotemporal process is straightforward:
find submachines with the symmetry tiling property. Reference \cite[Def.
43]{Shal01a} attempted a similar approach to define domains using \localstates:
the \emph{domain temporal phase} was defined as a strongly-connected set of
states where state transitions correspond to spatial transitions. A domain then
was a strongly-connected (in time) set of domain phases. Unfortunately, this can
be interpreted either as not allowing for single-phase domains, which are
prevalent, or else as allowing for nondomain submachines to be classified as
domain. In contrast, the symmetry tiling conditions in the above formulation
provide stricter conditions, in accordance with the symmetry group algebra, for
submachines to be classified as domain. For example, the simple cyclic symmetry 
groups for CA domains lead to cyclic domain submachines. Our formulation also
allows for a simpler (and more scalable) analysis through causal filtering.

\subsection{Structures as domain deviations}

With domain regions and their symmetries established, we now define coherent
structures in spatiotemporal systems as spatially localized, temporally persistent 
broken symmetries. For clarity, the following definition is given for a single spatial 
dimension, but the generalization to arbitrary spatial dimensions is straightforward.

\begin{defn}
A \emph{coherent structure} $\Gamma$ is a contiguous nondomain region
$\mathcal{R} \subset \lattice \otimes \mathbb{Z}$ of a spacetime field
$\stfield{}{}$ such that $\mathcal{R}$ has the following properties in the
semantic-filter fields of $\causalfield{}{} = \epsilon(\stfield{}{})$ or $T =
\tau(\stfield{}{})$:
\begin{enumerate}
\item \emph{Spatial locality}: Given a spatial configuration $\stfield{t}{}$ at 
	time $t$, $\Gamma$ occupies the spatial region $\mathcal{R}_t = [i : j]$ if 
	$\causalfield{t}{i:j}$ is bounded by domain states on 
	its exterior and contains nondomain states on its interior, $\causalfield{t}{i-1}
	\in \domain$, $\causalfield{t}{i} \notin \domain$, $\causalfield{t}{j} \notin
	\domain$, and $\causalfield{t}{j+1} \in \domain$.
\item \emph{Lagrangian temporal persistence}: Given $\Gamma$ occupies 
	the localized spatial region $\mathcal{R}_t$ at time $t$, $\Gamma$ 
	persists to the next time step if there is a spatially localized set of 
	nondomain states in $\causalfield{}{}$ at time $t+1$ occupying a contiguous spatial region 
	$\mathcal{R}_{t+1}$ that is within the depth-$1$ future lightcone of $\mathcal{R}_t$. 
	That is, for every pair of coordinates $ (r,t) \in \mathcal{R}_t$ and
	$(r',t+1) \in \mathcal{R}_{t+1}$, $||\site' - \site|| \leq c$.

\end{enumerate}
\end{defn}

For simplicity and generality we gave coherent structure properties in terms of
\localstate fields. For CAs, to which the FME operator may be applied, the
\DPID transducer filter may similarly be used to identify coherent structures.
However, the condition for temporal persistence is less strict: the regions
$\mathcal{R}_{t+1}$ and $\mathcal{R}_t$, when given over $T$ rather 
than $\causalfield{}{}$, must have finite overlap. That is,
there exists at least one pair of coordinates $(r,t) \in \mathcal{R}_t$ and
$(r', t+1) \in \mathcal{R}_{t+1}$ such that $r = r'$. Coherent
structures in CAs identified in this way are \DPID particles. Both notions 
of temporal persistence are referred to as Lagrangian since they allow 
$\Gamma$ to move through space over time.

Since \localstates are assigned to each point in spacetime, coherent structures
of all possible sizes can be described. The smallest scale possible is a single
spacetime point and the structure is captured by a single \localstate.
Larger structures are given as a set of states
localized at the corresponding spatial scale. Such sets may be arbitrarily
large and have (almost) arbitrary shape. In this way, the \localstates allow us
to discover complex structures, without imposing external templates on the
structures they describe. This leaves open the possibility of discovering novel
structures that are not readily apparent from a raw spacetime field or do not
fit into known shape templates.

\begin{figure*}[thp]
\centering
\includegraphics[width=\textwidth]{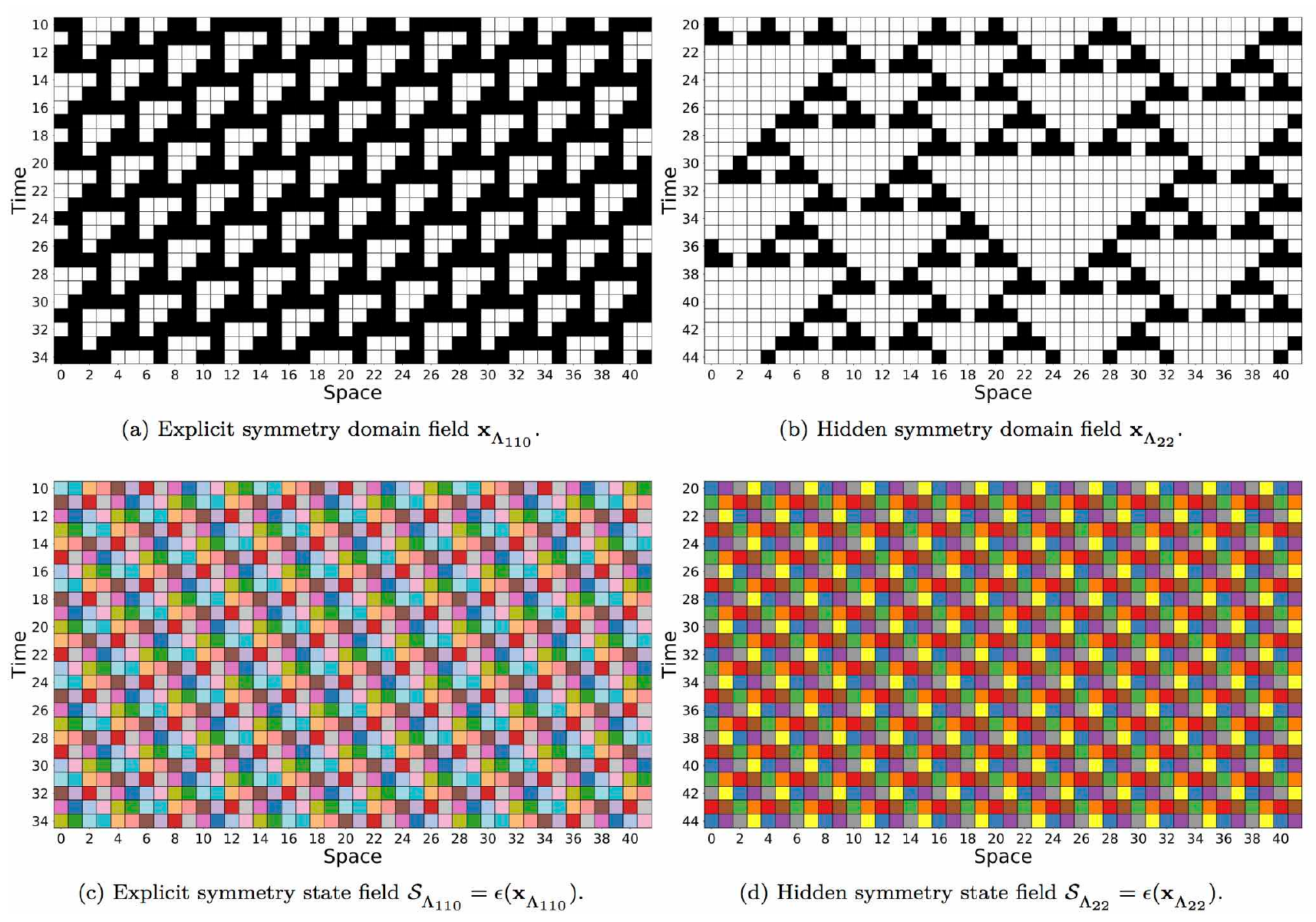}
\caption{Pure domain spacetime fields for explicit symmetry and hidden symmetry
	domains shown in (a) and (b) for ECA 110 and ECA 22, respectively.
	Associated \localstate fields fully display these symmetries in (c) and
	(d), with each unique color corresponding to a unique \localstate.
	For ECA 110, lightcone horizons $\phorizon = \fhorizon = 3$ were used
	and for rule 22 $\phorizon = 10$ and $\fhorizon = 4$.
	}
\label{symmetries}
\end{figure*}

\section{CA Structures}
\label{sec:CAStructures}

We now apply the theory of domains and coherent structures to discover patterns
in the spacetime fields generated by elementary cellular automata. We first
classify ECA domain types. For each class we analyze one exemplar ECA in
detail.  We begin describing the ECA's domain(s) and coherent structures
generated by the ECA, from both the \DPID and \localstate perspectives.

The analysis of domains and structures gives a sense of the correspondence
between \DPID and the \localstates. The general correspondence, found
empirically, between their descriptions of CA domains is as follows. For every
known \DPID CA domain language, a configuration from the language is used as an
initial condition to generate a pure domain field $\stfield{\domain}{}$. We see
that the spacetime shift operators over $\causalfield{\domain}{} =
\epsilon(\stfield{\domain}{})$ form symmetry groups with the same spatial
period, temporal period, and recurrence time as the \DPID domain language.

Though the CA dynamic $\Phi$ is not directly used to infer \localstates, the
correspondence between \DPID and \localstate domains shows that \localstates
incorporate detailed dynamical features and they can be used to discover
patterns and structures that can be defined directly from $\Phi$ using \DPID.

\subsection{CA domains and their classification}

ECA domains fall into one of two categories: \emph{explicit symmetry} or \emph{hidden symmetry}. In the \localstate formulation, a domain $\domain$ has explicit symmetry if the space and time shift operators, $\sigma_p$ and $\{\sigma^{s_n}\}$, which generate the domain symmetry group over $\causalfield{\domain}{} = \epsilon(\stfield{\domain}{})$, also generate that same symmetry group over $\stfield{\domain}{}$. That is, $\sigma_p (\stfield{\domain}{}) = \stfield{\domain}{}$ and $\sigma^{s_n}(\stfield{\domain}{} ) = \stfield{\domain}{}$, for all $n$. From this, we can see that explicit symmetry domains are periodic orbits of the CA, with the domain period equal to the orbit period. This follows since time shifts of the CA spacetime field are essentially equivalent to applying the CA dynamic $\Phi$; $\stfield{t+p}{} = \sigma_p (\stfield{}{})_t$ and $\stfield{t+p}{} = \Phi^p (\stfield{t}{})$. Thus, let $\state_\domain$ be any spatial configuration of a domain spacetime field, $\state_\domain = (\stfield{\domain}{})_t$, for any $t$, then $\Phi^p (\state_\domain) = \state_\domain$ if and only if $\sigma_p (\stfield{\domain}{}) = \stfield{\domain}{}$.

A hidden symmetry domain is one for which the time and space shift operators,
that generate the domain symmetry group over $\causalfield{\domain}{}$, do not
generate a symmetry group over $\stfield{\domain}{}$: $\sigma_p
(\stfield{\domain}{}) \neq \stfield{\domain}{}$ or
$\sigma^{s_n}(\stfield{\domain}{} ) \neq \stfield{\domain}{}$ or both.

In the \DPID formulation, a domain is classified as having explicit or hidden
symmetry based on the algebra of the domain languages. In this, group elements
are the strings of the spatial languages of the domain and the group action is
concatenation of the strings. If this algebra for every domain phase
$\domain_i$ is a proper group, $\domain$ has explicit symmetry. Otherwise, if
the algebra is something more general, like a semigroup or monoid, $\domain$
has hidden symmetry. Notably, hidden symmetry domains are associated with a
level of stochasticity in the raw spacetime field. We sometimes refer to these
as \emph{stochastic domains}. As the above domain algebra is only used for
classification here, we will not give the explicit mathematics. See Refs.
\cite[App. D]{Shal98a} or \cite{Kitc86a} for those details.

\begin{figure*}[htp]
  \centering
\includegraphics[width=\textwidth]{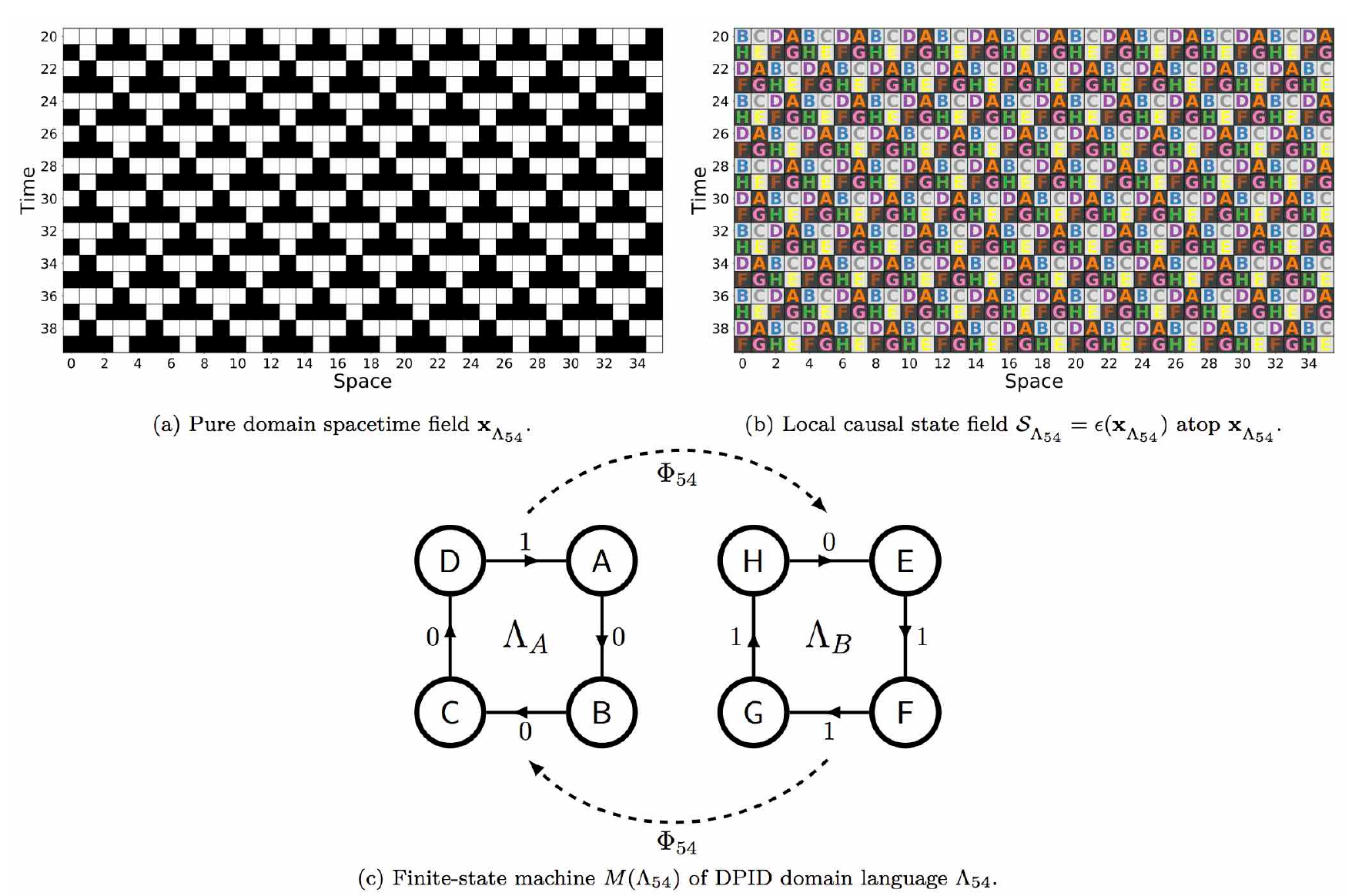}
\caption{ECA 54 domain: A sample pure domain spacetime field
	$\stfield{\domain}{}$ is shown in (a). This field is repeated with the
	associated local causal states $\causalfield{\domain}{} =
	\epsilon(\stfield{\domain}{})$ added in (b). Lightcone horizons $\phorizon = 
	\fhorizon = 3$ were used. The \DPID spacetime invariant
	set language is shown in (c). (Reprinted from Ref. \protect{\cite{Hans95a}}
	with permission.)
	}
\label{domain54}
\end{figure*}

Example domains from each category are shown in Figure~\ref{symmetries}. ECA
110 is given as the explicit symmetry example; a sample spacetime field
$\stfield{\domain_{110}}{}$ of its domain is shown in
Figure~\ref{symmetries}(a). The associated \localstate field
$\causalfield{\domain_{110}}{}$ is shown in Figure~\ref{symmetries}(c). Each
unique color corresponds to a unique \localstate. The \localstate field clearly
displays the domain's translation symmetries. ECA 110's domain has spatial
period $s = 14$ and temporal period $p = 7$. These are gleaned by direct
inspection of the spacetime diagram. Pick any color in
$\causalfield{\domain_{110}}{}$ and one must go through $13$ other colors
moving through space to return to the original color and, likewise, $6$ other
colors in time before returning. One can also see that at every time step
$\causalfield{\domain_{110}}{}$ has a single spatial tiling $w$ of the $14$
states. Thus, the recurrence time is $\widehat{p} = 1$. Finally, notice from
Figure~\ref{symmetries}(a) that spatial configurations of
$\stfield{\domain_{110}}{}$ are periodic orbits of $\Phi_{110}$, with orbit
period equal to the domain period, $p=7$. 

For a prototype hidden symmetry domain, ECA 22 is used. Crutchfield and McTague
used \DPID analysis to discover this ECA's domain in an unpublished
work \cite{Crut17a} that we used here to produce the domain spacetime
field $\stfield{\domain_{22}}{}$ shown in Figure~\ref{symmetries}(b). The
associated causal state field $\causalfield{\domain_{22}}{}$ is shown in
Figure~\ref{symmetries}(d). Unlike ECA 110's domain, it is not clear from
$\stfield{\domain_{22}}{}$ what the domain symmetries are. It is not even clear
there \emph{are} symmetries present from the raw spacetime field. However, the
causal state field $\causalfield{\domain_{22}}{}$ is immediately revealing.
Domain translation symmetries are clear. The domain is period $4$ in both space
and time: $p = s = 4$. There are eight unique \localstates in
$\causalfield{\domain_{22}}{}$ and, as the spatial period is $4$, the eight
states come in two distinct spatial tilings $w_1$ and $w_2$, each consisting of
$4$ states. And so, the recurrence time for ECA 22 is $\widehat{p} = 2$.
Shortly, we examine hidden symmetries in more detail to illustrate how the
\localstates lend a new semantics that exposes stochastic symmetries.

Having given concrete demonstrations of the new \localstate formulation of
domains and their classification in CAs, we move on to more detailed examples
that have been thoroughly studied from the \DPID perspective. In doing so, we
will see the strong correspondence between the two approaches, in terms of both
domains as well as the coherent structures which form atop the domains. 

\subsection{Explicit symmetries}

We start with a detailed look at ECA 54, whose domains and structures were
worked out in detail via \DPID \cite{Hans95a}. ECA 54 was said to support
``artificial particle physics'' and this emergent ``physics'' was specified by
the complete catalog of all its particles and their interactions. Here, we
analyze the domain and structures using \localstates and compare. Since the
particles (structures) are defined as deviations from a domain that has
explicit symmetries, the resulting higher-level particle dynamics themselves
are completely deterministic. As we will see later, this is not the case for
hidden symmetry systems; stochastic domains give rise to stochastic structures.

\subsubsection{ECA 54's domain}

A pure-domain spacetime field $\stfield{\domain}{}$ of ECA 54 is shown in
Fig.~\ref{domain54}(a). As can be seen, it has explicit symmetries and is
period $4$ in both time and space. From the \DPID perspective, though, it
consists of two distinct spatial-configuration languages, $\domain_A =
(0001)^*$ and $\domain_B = (1110)^*$, that map into each other under
$\Phi_{54}$; see Fig.~\ref{domain54}(c). This gives a recurrence time of 
$\widehat{p} = 2$. The finite-state machines,
$M(\Lambda_{A})$ and $M(\Lambda_{B})$, shown there for these languages each
have four states, reflecting the period-$4$ spatial translation symmetry:
$s = 4$. Although the domain's recurrence time  is $\widehat{p} = 2$,
the raw states $\stfield{t}{}$ are period $4$ in time due to a spatial phase
slip in the language evolution: $p = 4$. This is shown explicitly in
the spacetime machine given in Ref. \cite{Hans95a}. We can see that the machine
in Fig.~\ref{domain54}(c) fully describes the domain field in
Fig.~\ref{domain54}(a). At some time $t$, the system is either in $(0001)^*$ or
$(1110)^*$ and at the next time step $t+1$ it switches, then back again at
$t+2$, and so on.

Let's compare this with the \localstate analysis. The corresponding \localstate
field $\causalfield{\domain}{} = \epsilon(\stfield{\domain}{})$ was generated
from the pure domain field $\stfield{\domain}{}$ of Fig.~\ref{domain54}(a) via
causal filtering; see Fig.~\ref{domain54}(b). We reiterate here that this
reconstruction in no way relies upon the invariant set languages of
$\Lambda_{54}$ identified in \DPID. Yet we see that the \localstates correspond
exactly to $M(\Lambda_{54})$'s states. In total there are eight states, and
these appear as two distinct tilings in the field. These tilings correspond to
the two temporal phases of $\domain_{54}$: $w_A = [\sf{A}, \sf{B}, \sf{C},
\sf{D}] =$ $\domain_A$ and $w_B = [\sf{E}, \sf{F}, \sf{G}, \sf{H}] =$
$\domain_B$. At any given time $t$, a spatial configuration is tiled by only
one of these temporal phases, which each consist of $4$ states, giving a
spatial period $s= 4$. And, at the next time $t+1$ there are only states from
the other tiling.  Then back to previous tiling, and so, the evolution
continues. Thus, we can see the recurrence time is $\widehat{p} = 2$. In
contrast, the actual \localstates are temporally period $p = 4$, which is also
the orbit period of configurations in $\stfield{\domain}{}$, as can be seen in
Fig.~\ref{domain54}(a). This is in agreement with \DPID's invariant set
analysis, shown in Fig.~\ref{domain54}(c). As noted before and as will be
emphasized, there is a strong correspondence between \DPID's dynamically
invariant sets of spatially homogeneous configurations and the \localstate
description, both for coherent structures and the domains from which they are
defined.

\begin{figure}
\centering
\includegraphics{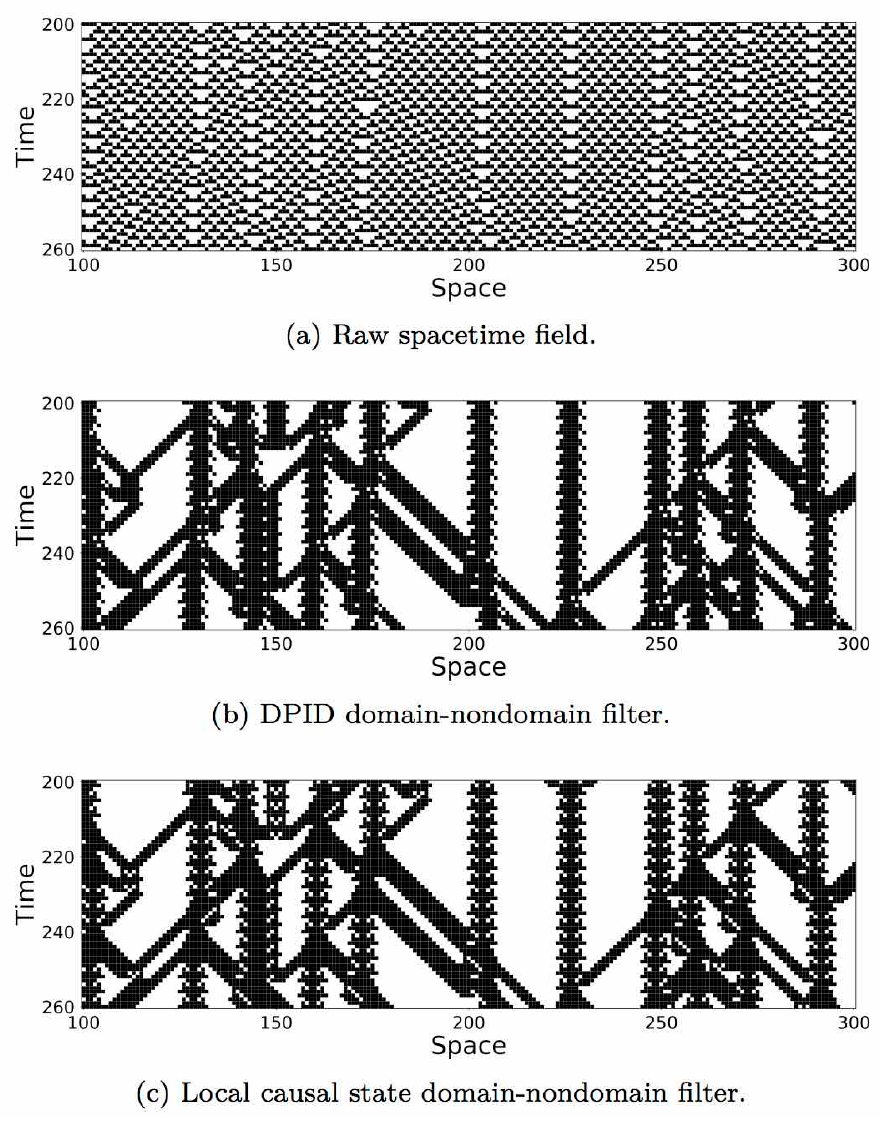}
\caption{Overview of ECA 54 structures: (a) A sample spacetime field evolved
	from a random initial configuration. (b) A filter that outputs white for
	cells participating in domains and black otherwise, using the \DPID
	definition of domain. (c) The analogous domain-nondomain filter that uses the
	\localstate definition of domain. Lightcone horizons $\phorizon = 
	\fhorizon = 3$ were used.
	}
\label{structures54}
\end{figure}

\subsubsection{ECA 54's structures}

Let's examine the structures (particles) supported by ECA 54 and their
interactions. Rule 54 organizes itself into domains and structures when started
with random initial conditions. A sample spacetime field $\stfield{}{}$
produced by evolving a random binary configuration under $\Phi_{54}$ is shown in
Fig.~\ref{structures54}(a). We first give a qualitative comparison of the
structures in this field from both the \DPID and \localstate perspectives.

From the \DPID side, a simple domain-nondomain filter is used with binary
outputs that flag sites in transducer filter field $T = \tau(\stfield{}{})$ as either domain (white) or
not domain (black).  Applying this filter to the spacetime field of
Fig.~\ref{structures54}(a) generates the diagram shown in
Fig.~\ref{structures54}(b).  Similarly, a domain-nondomain filter built from
\localstates when applied to Fig.~\ref{structures54}(a) gives the output shown
in Fig.~\ref{structures54}(c). For this filter, the eight domain \localstates
in $\causalfield{}{} = \epsilon(\stfield{}{})$ are in white and all other
\localstates black. While domain-nondomain detections differ site-by-site, we
see that in aggregate there is again strong agreement on the structures
identified by the two filter types.

There are four types of particles found in ECA 54 \cite{Hans95a}, which we can
now examine in detail. Before doing so, we must make a comment about the domain
transducer $\tau$ used by \DPID to identify structures.
As mentioned, a stack automaton is generally required, but may
be well-approximated with a finite-state transducer \cite{McTa04a}. A trade-off
is made with the transducer, however, since it must choose a direction to scan
configurations---left-to-right or right-to-left. To best capture the proper
spatial extent of a particle, an interpolation may be done by comparing right
and left scans. This was done in the domain-nondomain filter of
Fig.~\ref{structures54}(b). The bidirectional interpolation used does not
capture fine details of domain deviations. For the particle analysis that
follows, a single direction (left to right) scan is applied to produce each $T_t =
\tau(\stfield{t}{})$ in $T = \tau(\stfield{}{})$. A noticeable side-effect of the single direction scan is
that it covers only about half of any given particle's spatial extent.  (This
scan-direction issue simply does not arise in \localstate filtering.)

\begin{figure}
\centering
\includegraphics{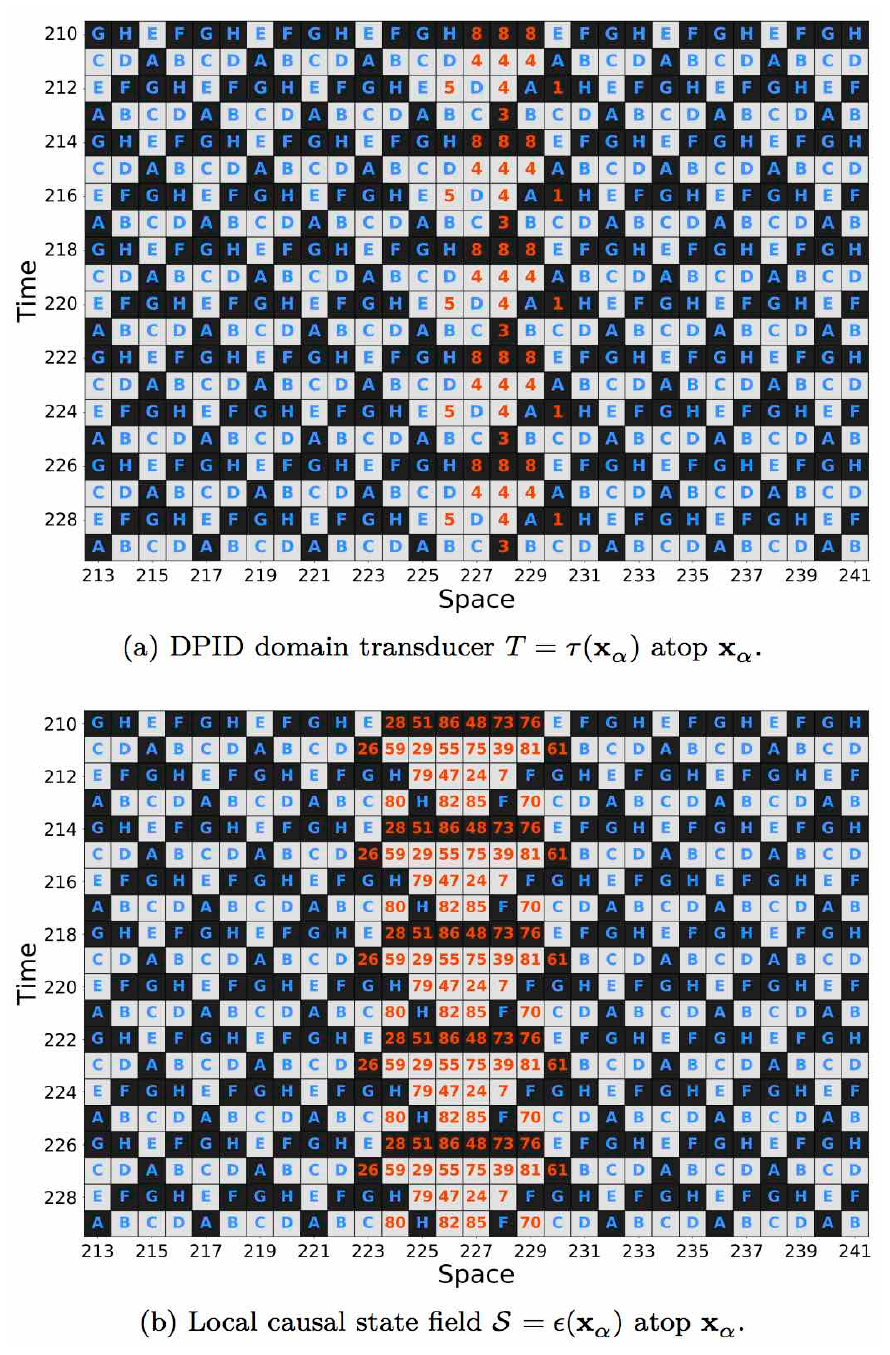}
\caption{ECA 54's $\alpha$ particle: In both (a) and (b) white ($0$) and black
	($1$) squares display the underlying ECA spacetime field
	$\stfield{\alpha}{}$. (a) The \DPID domain transducer filter $T =
	\tau(\stfield{\alpha}{})$ output is overlaid atop the spacetime field
	values of $\stfield{\alpha}{}$. Blue letters are sites participating in
	domain and red numbers are particular deviations from domain. (b)
	The \localstate field $\causalfield{}{} = \epsilon(\stfield{\alpha}{})$.
	The eight domain states are given by blue letters, all others by red numbers. In both
	diagrams, the non-domain sites outline the $\alpha$ particle of rule 54,
	according to the two different semantic filters. Lightcone 
	horizons $\phorizon = \fhorizon = 3$ were used.
 }
\label{alpha}
\end{figure}

The first structure we analyze is the large stationary $\alpha$ particle, shown
in Fig.~\ref{alpha}. For both diagrams the white and black squares represent
the values $0$ and $1$, respectively, of the underlying ECA field
$\stfield{\alpha}{}$. Overlaid blue letters and red numbers are the semantic
filter fields. In Fig.~\ref{alpha}(a) these come from the \DPID domain
transducer filtered field $T = \tau(\stfield{\alpha}{})$. In
Fig.~\ref{alpha}(b) they come from the \localstate field $\causalfield{}{} =
\epsilon(\stfield{\alpha}{})$.

For the \DPID domain transducer filtered field in Fig.~\ref{alpha}(a), overlaid
blue letters are sites flagged as participating in domain by the transducer
$\tau$, with the letter representing the spatial phase of the domain as given
by $M(\domain_{54})$. Red numbers correspond to sites flagged as various
deviations from domain \cite{Hans95a}. Here, the collection of such deviations
outlines the $\alpha$ particle's structure; though, as stated above, the
unidirectional transducer only identifies about half of the particle's spatial
extent. The main feature to notice is that the particle has a period-$4$
temporal oscillation. As the $\alpha$ is recognizable by eye from the raw field
values, one can see this period-$4$ structure is intrinsic to the raw spacetime
field and not an artifact of the domain transducer. However, the period-$4$
temporal structure is clearly displayed by the \DPID domain transducer
description of $\alpha$.

Figure~\ref{alpha}(b) displays the \localstate field $\causalfield{}{} =
\epsilon(\stfield{\alpha}{})$; the eight domain states are given as blue
letters, following Fig.~\ref{domain54}(b), and all other nondomain states,
which outline the $\alpha$, are red numbers. We see the \localstates fill out
the $\alpha$'s full spatial extent. Since the numeric labels for each state are
arbitrarily assigned during reconstruction, the $\alpha$'s spatial reflection
symmetry that is clearly present does not appear in the \localstate labels.
However, the underlying lightcones that populate the equivalence classes of
these states \emph{do} exhibit this symmetry. As with the \DPID domain
transducer description though, the \localstates properly capture the $\alpha$'s
temporal period-$4$.

We emphasize that coherent structures are behaviors of the underlying system
and, as such, they exist in the system's spacetime field. The semantic filter
fields are formal methods that identify sites in the underlying spacetime field
which participate in a particular structure. This is how overlay diagrams, like
Fig.~\ref{alpha}, derive their utility.

\begin{figure}
\includegraphics{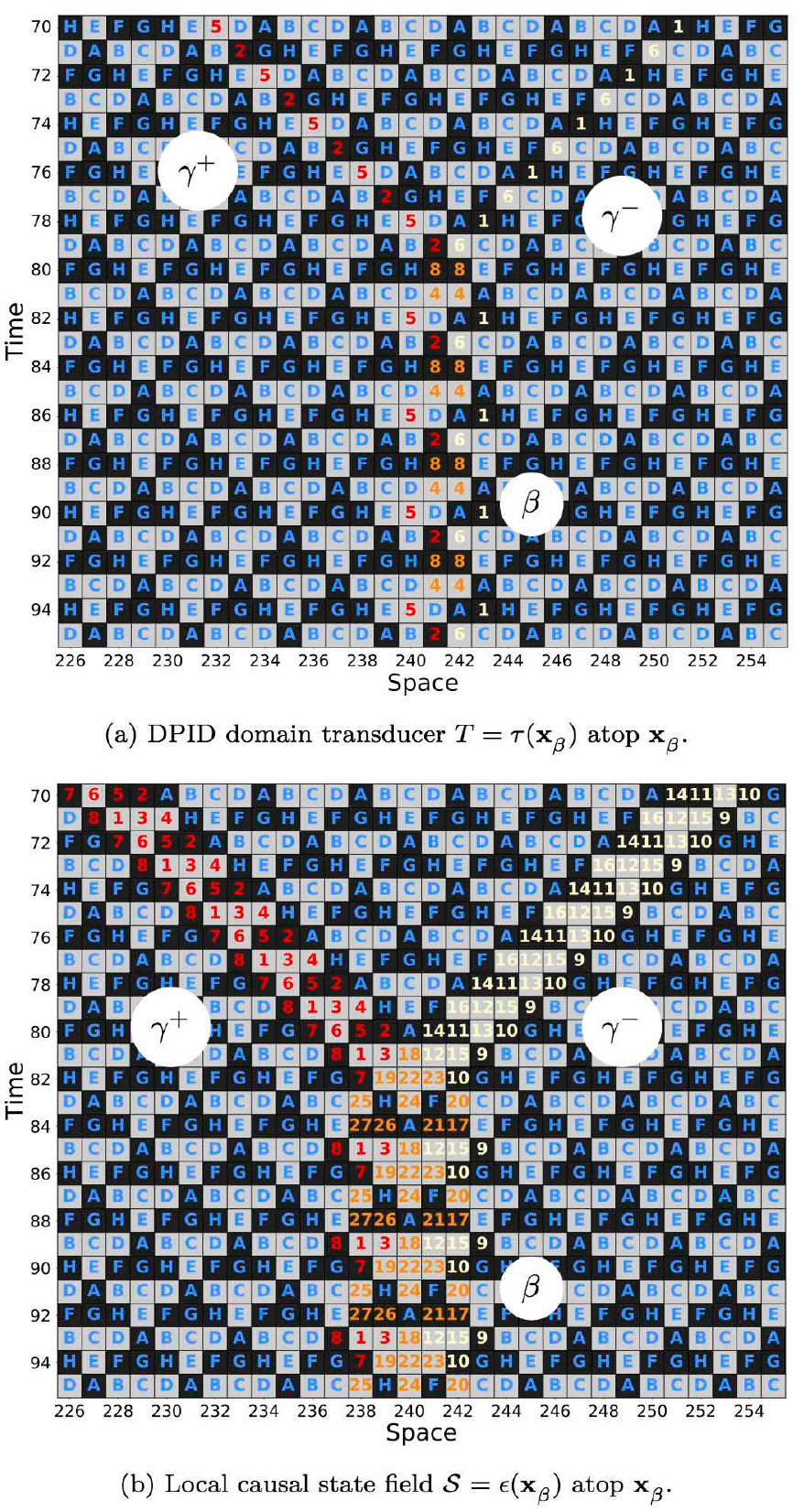}
\caption{ECA 54's $\gamma^+ + \gamma^- \rightarrow \beta$ interaction: In both
	diagrams the white ($0$) and black ($1$) squares display the underlying ECA
	spacetime field $\stfield{\beta}{}$. (a) The \DPID domain transducer filter
	$T = \tau(\stfield{\beta}{})$ output is overlaid atop the spacetime field
	values of $\stfield{\beta}{}$. Blue letters are sites identified by $T$ as
	participating in the domain. Colored numbers are sites identified
	as participating in one of the three remaining structures. The $\gamma^+$
	particle is outlined only by red numbers, $\gamma^-$ by yellow numbers, and
	$\beta$ by a combination of red, yellow, and orange. (b) The \localstate
	field $\causalfield{}{} = \epsilon(\stfield{\beta}{})$ is overlaid atop
	$\stfield{\beta}{}$. The eight domain states are in blue, and the other
	nondomain states are colored the same as in (a). Lightcone 
	horizons $\phorizon = \fhorizon = 3$ were used.
 }
\label{interaction}
\end{figure}

We discuss the three remaining structures of ECA 54 by examining an interaction
among them; the left-traveling $\gamma^-$ particle can collide with the
right-traveling $\gamma^+$ particle to form the $\beta$ particle. This
interaction is displayed with overlay diagrams in Fig.~\ref{interaction}. The
values of the underlying field $\stfield{\beta}{}$ are given by white ($0$) and
black ($1$) squares. The \DPID domain transducer filter field $T =
\tau(\stfield{\beta}{})$ is overlaid over top of $\stfield{\beta}{}$ in
Fig.~\ref{interaction}(a) and the \localstate field $\causalfield{}{} =
\epsilon(\stfield{\beta}{})$ atop $\stfield{\beta}{}$ in
Fig.~\ref{interaction}(b).

In both cases, the color scheme is as follows. Sites identified by the semantic
filters as participating in a domain are colored blue, with the letters
specifying the particular phase of the domain. In Fig.~\ref{interaction}(a) the
domain phases are specified by $T$ and in Fig.~\ref{interaction}(b) they are
specified by $S$.  And, as we saw in Fig.~\ref{domain54} and can see here,
these specifications of $\domain_{54}$ are identical. For both
Figs.~\ref{interaction}(a) and \ref{interaction}(b), nondomain sites
participating in the $\gamma^+$ are flagged with red, those participating in
the $\gamma^-$ with yellow, and those uniquely participating in the $\beta$
with orange.

As with the $\alpha$ particle, the \localstate description better covers the
particles' spatial extent, but both filters agree on the temporal oscillations
of each particle. Both $\gamma$s are period $2$ and $\beta$ is period $4$.
Unlike the $\alpha$ and $\beta$, the $\gamma$ particles are not readily
identifiable by eye. They arise as a result of a phase slip in the domain. For
example, a spatial configuration with a $\gamma$ present is of the form
$\domain_A \; \gamma \; \domain_B$.

Related to this, we point out here an observation about this interaction that
illustrates how our methods uncover structures in spatiotemporal systems.  At
the top of each diagram in Fig.~\ref{interaction} the spatial configurations
are of the form $\domain_A \; \gamma^+ \; \domain_B \; \gamma^- \; \domain_A$.
At each subsequent time step, the domains change phase $A \rightarrow B$ and $B
\rightarrow A$ and the intervening domain region shrinks as the $\gamma$s move
towards each other. The intervening domain disappears when the $\gamma$s
finally collide. Then we have local configurations of the form $\domain_A \;
\beta \; \domain_A$. However, there is an indication that a phase slip between
these domain regions still happens ``inside'' the $\beta$ particle. Notice in
Fig.~\ref{interaction} there are several spatial configurations (horizontal
time slices) in which domain states appear inside the $\beta$ that are the
opposite phase of the bordering domain phases, indicating a phase slip. Also,
the states constituting the $\gamma$s are found as constituents of the $\beta$.
For the \DPID domain transducer $\tau$, each $\gamma$ consists of just two
states, and all four of these states (two for each $\gamma$) are found in the
$\beta$. In the \localstate field $S$, each $\gamma$ is described by eight
\localstates. Not all of these show up as states of the $\beta$, but several
do. Those $\gamma$ states that do show up in the $\beta$ appear in the same
spatiotemporal configurations they have in the $\gamma$s.

These observations tell us about the underlying ECA's behavior and so
can be gleaned from the raw spacetime field itself. That said, the
discovery that the $\beta$ particle is a ``bound state'' of two $\gamma$s and
that it contains an internal phase slip of the bordering domain regions is
not at all obvious from inspecting raw spacetime fields. That is, $\gamma^+ \;
\domain \; \gamma^- \rightarrow \beta $. Such structural discovery, however,
is greatly facilitated by the coherent structure analysis. To emphasize, these
insights concern the intrinsic organization embedded in the spacetime fields
generated by the ECA. No structural assumptions, beyond the very basic
definitions of \localstates, are required.

Let's recapitulate the correspondence between the independent \DPID and
\localstate descriptions of the ECA 54 domain and structures. From the \DPID
perspective, the ECA 54 domain $\domain_{54}$ consists of two homogeneous
spatial phases that are mapped into each other by $\Phi_{54}$. In contrast,
$\domain_{54}$ is described by a set of \localstates with a spacetime
translation symmetry tiling. The two descriptions agree completely, giving a
spatial period $4$, temporal period $4$, and recurrence time of $2$. On
the one hand, for ECA $54$'s structures \DPID directly uses domain information
to construct a transducer filter $T = \tau(\stfield{}{})$ that
identifies structures as groupings of particular domain deviations. On the
other, the \localstates are assigned uniformly to spacetime field sites via
causal filtering $\causalfield{}{} = \epsilon(\stfield{}{})$. Domains and sites
participating in a domain are found by identifying spatiotemporal symmetries in
the \localstates. Coherent structures are then localized deviations from these
symmetries. Though the agreement is not exact as with the domain, \DPID and the
\localstates still agree to a large extent on their descriptions of ECA 54's
four particles and their interactions.

\begin{figure}
\centering
\includegraphics{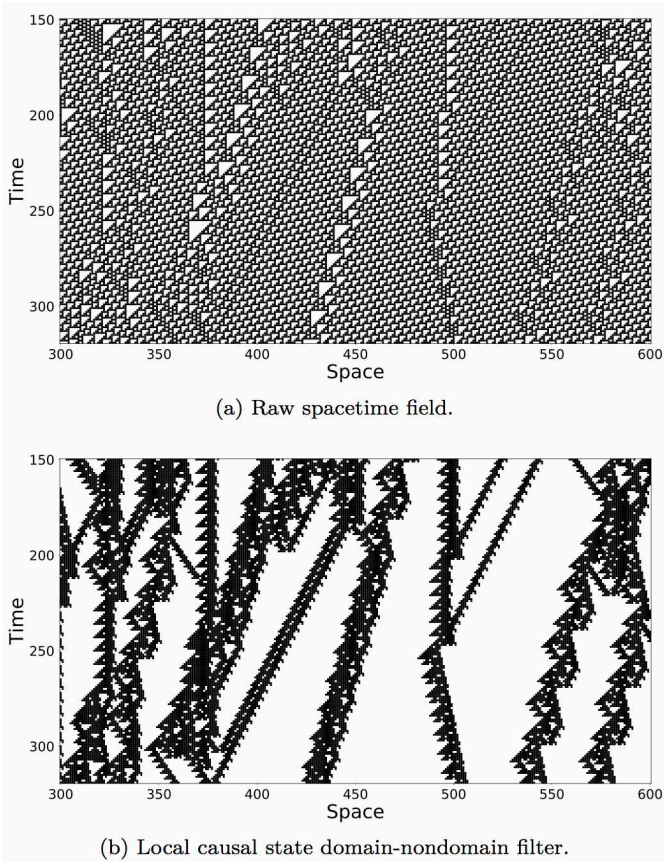}
\caption{ECA 110 structures: (a) A sample field evolved from a random initial
	configuration. (b) A \localstate domain-nondomain filter with domain sites
	in white and nondomain in black. Lightcone 
	horizons $\phorizon = \fhorizon = 3$ were used.
	}
\label{rule110}
\end{figure}

\subsubsection{ECA 110}

As the most complex explicit symmetry ECA, ECA 110 is worth a brief mention.
It is the only ECA proven to support universal computation (on a specific
subset of initial configurations) and implements this using a subset of the
ECA's coherent structures \cite{Cook04a}. This was shown by mapping ECA 110's
particles and their interactions onto a cyclic tag system that emulates a Post
tag system which, in turn, emulates a universal Turing machine. A
domain-nondomain filter reveals several of ECA 110's particles used in the
implementation; see Fig.~\ref{rule110}. The ECA 110 domain was displayed in
Figs.~\ref{symmetries}(a) and \ref{symmetries}(c), as the example for explicit
symmetry domains. The domain has a single phase, rather than two phases like
ECA 54's, and requires $14$ states, as opposed to ECA 54's combined $8$. The
ECA 110's highly complex behavior surely derives from the heightened complexity
of its domain. Exactly how, though, remains an open problem.

\subsection{Hidden stochastic symmetries}

Our attention now turns to ECAs with hidden symmetries and stochastic domains.
These are the so-called ``chaotic'' ECAs. Since the structure of an ECA's
domain heavily dictates the overall behavior, stochastic domains give rise to
stochastic structures and hence, in combination, to an overall stochastic
behavior. To be clear, since all ECA dynamics are globally deterministic---the
evolution of spatial configurations is deterministic---the stochasticity here
refers to local structures rather than global configurations. In contrast to
explicit symmetry ECAs whose structures are largely identifiable from the raw
spacetime field, the structures found in stochastic-domain ECAs are often not
at all apparent. In this case the ability of our methods to facilitate the
discovery and description of such hidden structures is all the more important
and sometimes even necessary. While the distinction between stochastic and
explicit symmetry domains does not make a difference when determining \DPID's
spacetime invariant sets, \localstate inference is relatively more difficult
with stochastic domains, usually requiring large lightcone depths and an
involved domain-structure analysis.

Here, we examine ECA 18 in detail, as its stochastic domain is relatively
simple and well understood. An empirical domain-structure analysis of ECA 18
was first given in Ref. \cite{Gras83a} and then more formally in Refs.
\cite{Lind84a,Elor91a,Bocc91a,Elor94a}, which notes the domain's temporal
invariance. It was not until the FME was introduced in Ref. \cite{Hans90a} that
this was rigorously proven and shown to follow within the more \DPID general
framework. The distinguishing feature of ECA 18's domain observed in the early
empirical analysis was that the lookup table $\phi_{18}$ becomes
\emph{additive} when restricted to domain configurations. Specifically, when
restricted to domain, $\phi_{18}$ is equivalent to $\phi_{90}$, which is the
sum mod $2$ of the outer two bits of the local neighborhood; $\stfield{t+1}{r}
= \phi_{90} (\stfield{t}{r-1} \stfield{t}{r} \stfield{t}{r+1}) =
\stfield{t}{r-1} + \stfield{t}{r+1} \; (\mathrm{mod} \; 2)$.

ECA 18's structures illustrate
additional complications of \localstate analysis with stochastic symmetry
systems. Nondomain states of ECA 54 and other explicit symmetry ECAs always
indicate a particle or particle interaction, after transients. This is not the
case with chaotic ECAs, and our formal definition is needed to identify ECA
18's coherent structures.

\begin{figure}
\centering
\includegraphics{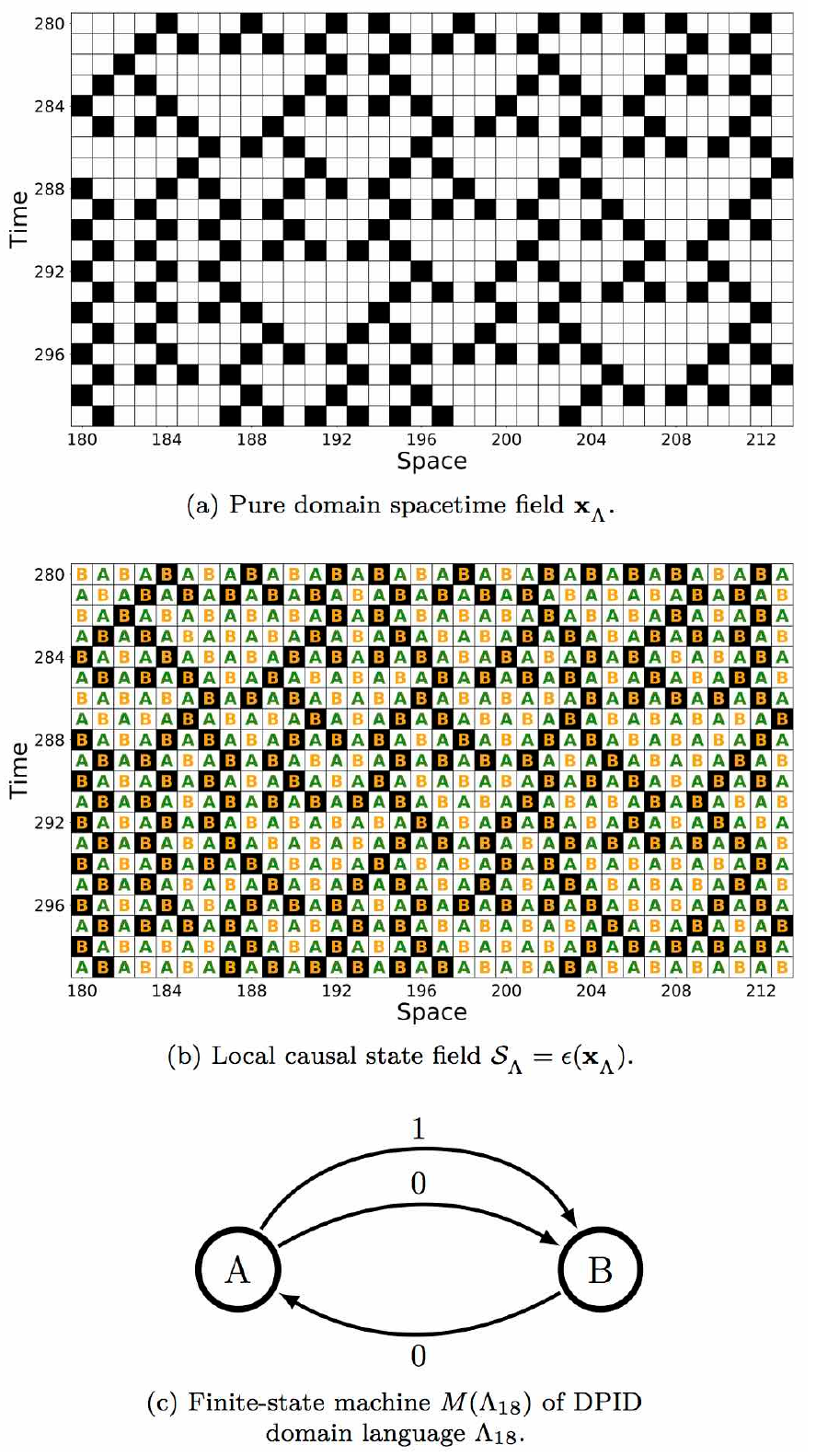}
\caption{ECA 18 domain: (a) Iterates of a sample pure domain spacetime field
	$\stfield{\domain}{}$, white and black are values $0$ and $1$,
	respectively. (b) The same domain field with the \localstate field
	$\causalfield{\domain}{} = \epsilon(\stfield{\domain}{})$ overlaid. 
	Lightcone horizons $\phorizon = 8$ and $\fhorizon = 3$ were used. (c) The
	finite-state machine $M(\domain_{18})$ of the \DPID invariant set language
	of the ECA 18 domain $\domain_{18}$. (Reprinted with permission from Ref \cite{Hans90a}.)
	}
\label{domain18}
\end{figure}

\subsubsection{ECA 18's domain}

Iterates of a pure domain spacetime field $\stfield{\domain_{18}}{}$ for the
ECA 18 domain $\domain_{18}$ is shown in Fig.~\ref{domain18}(a). White and
black cells represent site values $0$ and $1$, respectively. A symmetry is not
apparent in the spacetime field. One noticeable pattern, though, is that $1$s
(black cells) always appear in isolation, surrounded by $0$s on all four sides.
This still does not reveal symmetry, since neither time nor space shifts match
the original field. When scanning along one dimension, making either timelike
or spacelike moves (vertically or horizontally), one sees that every other site
is always a $0$ and the sites in between are \emph{wildcards}---they can be
either $0$ or $1$. Making this identification finally reveals the symmetry in
the ECA 18 domain \cite{Hans90a}.

In contrast to this ad hoc description, the $0$-wildcard pattern is clearly and
immediately identified in the \localstate field $\causalfield{\domain}{} =
\epsilon(\stfield{\domain}{})$, shown in Figure~\ref{domain18}(b). State $A$
occurs on the fixed-$0$ sites and state $B$ on the wildcard sites. And, these
states occur in a checkerboard symmetry that tiles the spacetime field. An
interesting observation of this symmetry group is that it has rotational
symmetry, in addition to the time and space translation symmetries. This is a
rotation, though, in spacetime. While unintuitive at first, the above discussion
shows this spacetime rotational symmetry is not just a coincidence. The
$0$-wildcard semantics applies for \emph{both} spacelike and timelike scans
through the field.

The \DPID invariant-set language for this domain is given in
Figure~\ref{domain18}(c). Not surprisingly, this is the $0$-wildcard language.
It is easy to see that $\phi_{18}$ creates a tiling of $0$-wildcard local
configurations. Also, note the transition branching (the wildcard) leaving
state $A$ indicates a semigroup algebra. This identifies $\domain_{18}$ as a
stochastic symmetry domain. We again see a clear correspondence between the
\localstate identification of the domain and that of \DPID. Both give spatial period 
$s = 2$, temporal period $p = 2$, and recurrence time $\widehat{p} = 1$, as there 
is a single \localstate tiling and a single \DPID spatial language, both corresponding to the 
$0$-wildcard pattern. 

\begin{figure}
\centering
\includegraphics{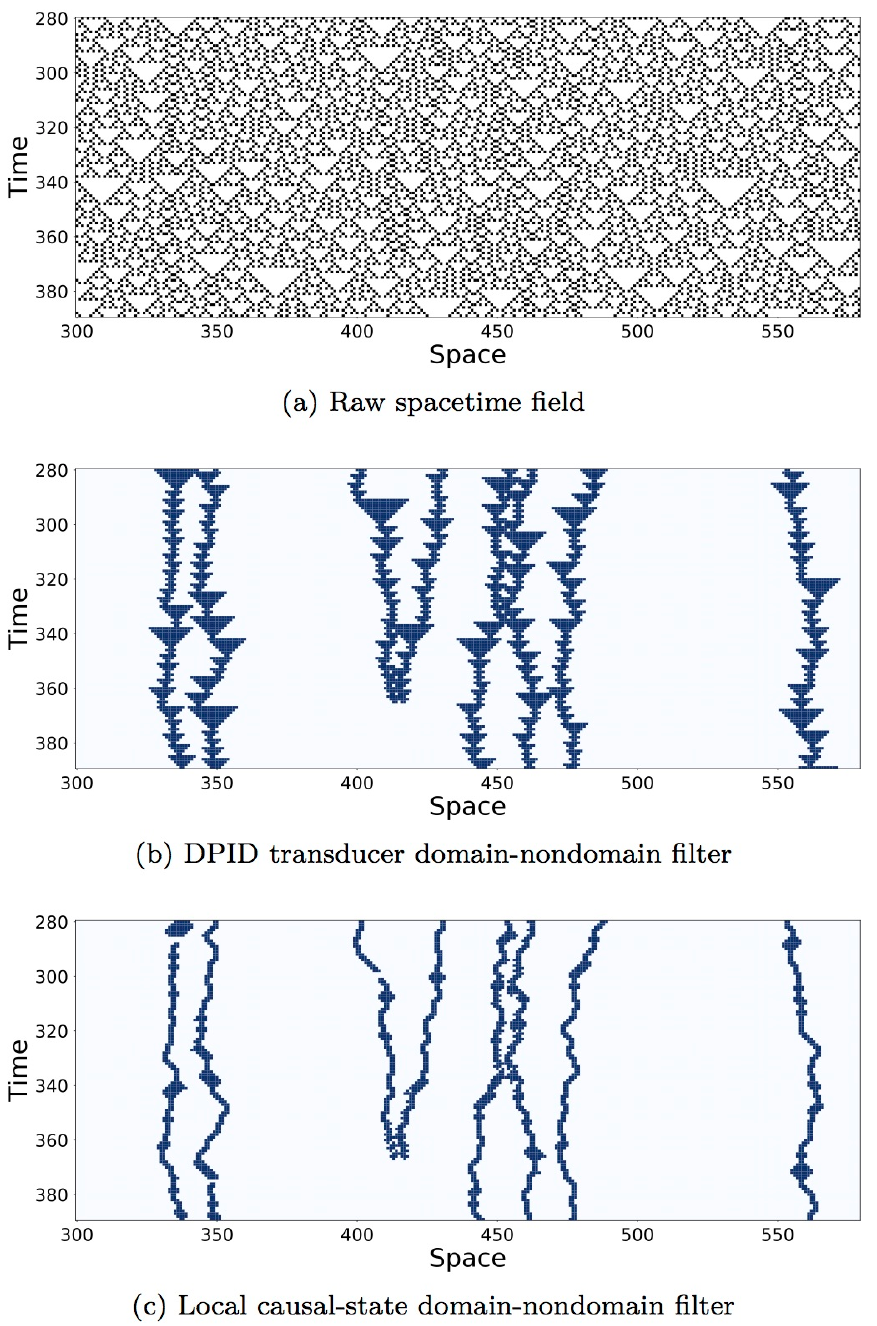}
\caption{ECA 18 structures: (a) Sample spacetime field evolved under ECA 18
	from a random initial configuration. (b) Spacetime field after filtering
	with domain regions in white and coherent structures in blue, using the
	\DPID domain transducer. (c) Spacetime field filtered with domain regions
	in white and structures in blue, using \localstates. The occasional gap in
	the structures is an artifact of using finite-depth lightcones during
	reconstruction of local causal states. Lightcone horizons 
	$\phorizon = 8$ and $\fhorizon = 3$ were used.
	}
\label{rule18}
\end{figure}

\subsubsection{ECA 18's structures}

ECA 18's two-state domain $\domain_{18}$ supports a single type of coherent
structure---the $\alpha$ particle that appears as a phase-slip in the spatial
period-$2$ domain and consists of local configurations $10^{2k}1$, $k = 0, 1, 2,
\ldots$.  The domain's stochastic nature drives the $\alpha$s in an unbiased
left-right random-walk. When two collide they pairwise annihilate; resolving
each $\alpha$'s spatial phase shift. To clarify, the $\alpha$ of ECA 18 has
no relation to the $\alpha$ of ECA 54.

Figure~\ref{rule18} shows these structures as they evolve from a random initial
configuration under $\Phi_{18}$. The raw spacetime field is given in
Fig.~\ref{rule18}(a) with the \DPID transducer domain-nondomain filter
(bidirectional scan interpolation) in Fig.~\ref{rule18}(b) and the \localstate
domain-nondomain filter in Fig.~\ref{rule18}(c). With the aid of these domain
filters, visual inspection shows that ECA 18's structures are, in fact,
pairwise annihilating random-walking particles. This was explored in detail by
Ref. \cite{Crut92a}.

As noted above, the domain-structure \localstate analysis for stochastic domain
systems is generally more subtle. In the \DPID analysis, ECA 18 consists solely
of the single domain and random-walking $\alpha$ particle structures. Thus,
using the \DPID transducer to filter out sites participating in domains leaves
only $\alpha$ particles, as done in Fig.~\ref{rule18}(b). The situation is more
complicated in the \localstate analysis. As described in more detail shortly,
filtering out domain states leaves behind more than the structures. Why exactly
this happens is the subject of future work. The field shown in
Fig.~\ref{rule18}(c) was produced from a coherent structure filter, rather than
from a domain-nondomain filter. There, \localstates that fit the coherent
structure criteria are colored blue and all others are colored white.

To illustrate the more involved \localstate analysis let's take a closer look
at the $\alpha$ particle. This also highlights a major difference between \DPID
and \localstate analyses. As the \DPID transducer is strictly a spatial
description it can identify structures that grow in a single time step to
arbitrary size. One artifact of this is that the spatial growth can exceed the
speed of local information propagation and thus make structures appear acausal.
The \localstates, however, are constructed from lightcones and so naturally
take into account this notion of causality. They cannot describe such acausal
structures. Accounting for this, though, there is a strong agreement between
the two descriptions.

\begin{figure*}
\vspace{-8mm}
\centering
\includegraphics{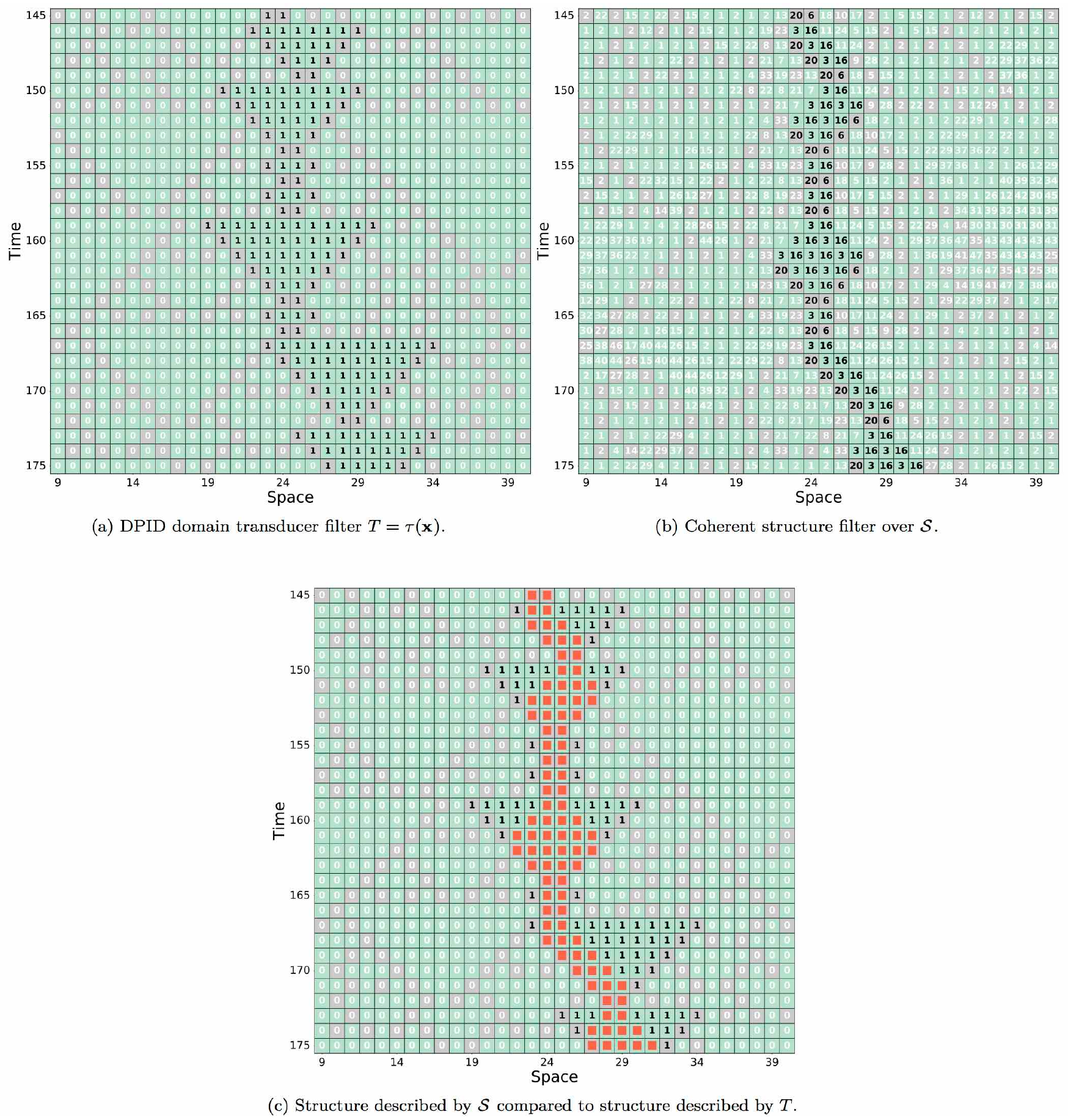}
\caption{Comparative analysis of ECA 18's $\alpha$ particle: In all three
	spacetime diagrams, the underlying ECA field values of $0$ and $1$ are
	represented as green and gray squares, respectively.
	(a) \DPID domain transducer filtered field $T = \tau(\stfield{}{})$ with
	bidirectional scan interpolation. Domain sites are identified with white
	$0$s and particle sites with black $1$s.  
	(b) A coherent structure  causal filter; \localstate field
	$\causalfield{}{} = \epsilon(\stfield{}{})$ with nondomain \localstates
	states satisfying the coherent structure definition are colored black with
	all other states colored white. Lightcone horizons 
	$\phorizon = 8$ and $\fhorizon = 3$ were used.
	(c) Comparison of the structures from the two methods: The \DPID transducer
	filter of (a) with sites that have \localstates identified as the coherent
	structure in (b) given a red square label.
	}
\label{particle18}
\end{figure*}

From the perspective of the \DPID domain transducer $\tau$
ECA 18's $\alpha$ particles are simple to understand. From the domain language
in Fig.~\ref{domain18}(c), the domain-forbidden words are those in the regular
expression $1(00)^*1$. That is, pairs of ones with an even number of $0$s
(including no $0$s) in between. This is the description of $\alpha$ particles
at the spatial configuration level. The \DPID bidirectional scan interpolation
domain transducer perfectly captures $\alpha$ described this way; see
Fig.~\ref{particle18}(a). To aid in visual identification we employed a
different color scheme for Fig.~\ref{particle18}: the underlying ECA field
values are given by green ($0$) and gray ($1$) squares. For the \DPID
transducer filtered field $T = \tau(\stfield{}{})$ in Fig.~\ref{particle18}(a), overlaid white $0$s
identify domain sites and black $1$s identify particle sites. Every local
configuration identified as an $\alpha$ is of the form $1(00)^*1$. As noted
above, however, $\alpha$s described in this way can grow in size arbitrarily in
a single time step as the number of pairs of zeros in $1(00)^*1$ is unbounded.

\Localstate inference---whether topological \cite{Rupe17c} or probabilistic
\cite{Shal03a}---is \emph{unsupervised} in the sense that it uses only raw
spacetime field data and no other external information such as the CA rule used
to create that spacetime data. Once states are inferred, further steps are
needed for coherent structure analysis.

The first step is to identify domain states in the \localstate field
$\causalfield{}{} = \epsilon(\stfield{}{})$. They tile spacetime regions, i.e.,
domain regions.
For explicit-symmetry domain ECAs this step is sufficient for creating a domain-structure filter. Tiled domain states can be easily identified and all other states outline ECA structures or their interactions.
The situation is more subtle, however, for ECAs with stochastic domains.
A detailed description of the implementation of additional ``decontamination'' steps is given in a sequel.

For our purposes here, though, it suffices to strictly apply the definition of
coherent structures after this first ``out of the box'' unsupervised causal
filter. The initial unsupervised filtered spacetime diagram
identifies a core set of states that are spatially localized and temporally
persistent. A coherent structure filter then isolates these states by coloring
them black and all other states white in the \localstate field
$\causalfield{}{}$. The output of this filter is shown in
Fig.~\ref{particle18}(b). The growth rate of the structures identified in this
way---by the \localstates---is limited by the speed of information propagation,
which for ECAs is unity. Applying this growth-rate constraint on the \DPID
structure transducer, one again finds strong agreement. A comparison is shown
in Fig.~\ref{particle18}(c). It shows the output of the \DPID filter applied to
the spacetime field of Fig.~\ref{particle18}(a) and, in red, sites
corresponding to the structure according to the \localstates in
Fig.~\ref{particle18}(b).

\section{Discussion}

Having laid out our coherent structure theory and illustrating it in some
detail, it is worth looking back, as there are subtleties worth highlighting.
The first is our use of the notion of \emph{semantics}, which derives from the
\emph{measurement semantics} introduced in Ref. \cite{Crut91b}. Performing
causal filtering $\causalfield{}{} = \epsilon(\stfield{}{})$ may at first seem
counterproductive, especially for binary fields like those generated by ECAs,
as the state space of the system is generally \emph{larger} in
$\causalfield{}{}$ than in $\stfield{}{}$. As the local state space of ECAs is
binary, complexity is manifest in how the sites interact and arrange
themselves.  Not all sites in the field play the same role. For instance, in
ECA 110's domain, Fig.~\ref{symmetries}(a), the $0$s in the field
group together to form a triangular shape. This triangle has a bottom-most $0$
and a rightmost $0$, but they are both still $0$s. To capture the
semantics of ``bottom-most'' and ``rightmost'' $0$ of that triangle shape, a
larger local state space is needed. And, indeed, this is exactly the manner in
which the \localstates capture the semantics of the underlying field. We saw a
similar example with the fixed-$0$ and wildcard semantics of $\domain_{18}$.

The values in the fields $\causalfield{}{} = \epsilon(\stfield{}{})$ and $T =
\tau(\stfield{}{})$ are not measures of some quantity, but rather semantic
labels. For the \localstates, they are labels of equivalence classes of local
dynamical behaviors. For the \DPID domain transducer, they label sites as being
consistent with the domain language $\domain$ or else as the particular manner
in which they deviate from that language.

This, however, is only the first level of semantics used in our coherent
structure theory. While the filtered fields $\causalfield{}{} =
\epsilon(\stfield{}{})$ and $T = \tau(\stfield{}{})$ capture semantics of the
original field $\stfield{}{}$, to identify coherent structures a new level of
semantics on top of these filtered fields is needed. These are semantics that
identify sites as domain or coherent structure using $\causalfield{}{}$ and
$T$. For the \DPID domain transducer $T$, the domain semantics are by
construction built into $T$. Our coherent structure definition adds the
necessary semantics to identify collections of nondomain sites as participating
in a coherent structure.

For the \localstates, one may think of the field $\causalfield {}{} =
\epsilon(\stfield{}{})$ as being the semigroup level of semantics. That is,
they represent pattern and structure as generalized symmetries of the
underlying field $\stfield{}{}$. This is the same manner in which the \eM
captures pattern and structure of a stochastic process with semigroup algebra
\cite{Crut91b}. The next level of semantics, used to identify domains, requires
finding explicit symmetries in $\causalfield{}{}$. Thus, domain semantics are
the group-theoretic level of semantics, since domains are identified by
spacetime translation symmetry groups over $\causalfield{}{}$. With states
participating in those symmetry groups identified, our coherent structure
definition again provides the necessary semantics to identify structures in
$\stfield{}{}$ through $\causalfield{}{}$. These remarks hopefully also clarify
the interplay between group and semigroup algebras in our development.

Lastly, we highlight the distinction between a CA's local update rule $\phi$
and its global update $\Phi$---the CA's equations of motion. For many CAs, as
with ECAs, $\Phi$ is constructed from simultaneous synchronous application of
$\phi$ across the lattice. In a sense, then, there is a simple relation between
$\phi$ and $\Phi$. However, as demonstrated by many ECAs, most notably the
Turing complete ECA 110, the behaviors generated by $\Phi$ can be
extraordinarily complicated, even though $\phi$ is extraordinarily simple. This
is why complex behaviors and structures generated by ECAs are said to be
\emph{emergent}.

This point is worth emphasizing here due to the relationship between past
lightcones and $\phi^i$ for CAs. Since the \localstates are equivalence classes
of past lightcones, they are equivalence classes of the elements of $\phi^i$
for CAs. Thus, the system's local dynamic is directly embedded in the
\localstates. As we saw, the \localstates are capable of capturing emergent
behaviors and structures of CAs and so, in a concrete way, they provide a
bridge between the simple local dynamic $\phi$ and the emergent complexity
generated by $\Phi$. Moreover, the correspondence between the \localstate and
\DPID domain-structure analysis shows the particular equivalence relation over
the elements of $\phi^i$ used by the \localstates captures key dynamical
features of $\Phi$, used explicitly by \DPID.

The relationship, though, between $\phi^i$ and $\Phi$ captured by the
\localstates is not entirely transparent, as most clearly evidenced by the need
for behavior-driven reconstruction of the \localstates. Given a CA lookup table
$\phi$, one may pick a finite depth $i$ for the past lightcones and easily
construct $\phi^i$.  It is not at all clear, however, how to use $\Phi$ to
generate the equivalence classes over the past lightcones of $\phi^i$ that have
the same conditional distributions over future lightcones. The only known way
to do this is by brute-force simulation and reconstruction, letting $\Phi$
generate past lightcone-future lightcone pairs directly.

\section{Conclusions}

Two distinct, but closely related, approaches to spatiotemporal computational
mechanics were reviewed: \DPID and \localstates. From them, we developed a
theory of coherent structures in fully discrete dynamical field theories. Both
approaches identify special symmetry regions of a system's spatiotemporal
behavior---a system's domains. We then defined coherent structures as localized
deviations from domains; i.e., coherent structures are locally broken domain
symmetries.

The \DPID approach defines domains as sets of homogeneous spatial
configurations that are temporally invariant under the system dynamic. In 1+1
dimension systems, dynamically important configuration sets can be specified as
particular types of regular language. Once these domain patterns are
identified, a domain transducer $\tau$ can be constructed
that filters spatial configurations $T_t = \tau(\stfield{t}{})$, identifying sites that participate in
domain regions or that are the unique deviations from domains. Finding a
system's domains and then constructing domain transducers requires much
computational overhead, but full automation has been demonstrated. Once
acquired, the domain transducers provide a powerful tool for analyzing emergent
structures in discrete, deterministic 1+1 dimension systems. The theory of
domains as dynamically invariant homogeneous spatial configurations is easily
generalizable beyond this setting, but practical calculation of configuration
invariant sets in more generalized settings presents enormous challenges.

The \localstate approach, in contrast, generalizes well. Both in theory and in
practice, under a caveat of computational resource scaling. It is a more direct
generalization of computational mechanics from its original temporal setting.
The causal equivalence relation over pasts based on predictions of the future
is the core feature of computational mechanics from which the generalization
follows. \Localstates are built from a local causal equivalence relation over
past lightcones based on predictions of future lightcones. \Localstates provide
the same powerful tools of domain transducers, and more. Being equivalence
classes of past lightcones, which in the deterministic setting are the system's
underlying local dynamic, \localstates offer a bridge between emergent structures and
the underlying dynamic that generates them.

In both, patterns and structures are discovered rather than simply recognized.
No external bias or template is imposed, and structures at all scales may be
uniformly captured and represented. These representations greatly facilitate
insight into the behavior of a system, insights that are intrinsic to a system
and are not artifacts of an analyst's preferred descriptional framework. ECA
54's $\gamma^+ + \domain + \gamma^- \rightarrow \beta$ interaction exemplifies this.

\DPID domain transducers utilize full knowledge of a system's underlying
dynamic and, thus, perfectly capture domains and structures. \Localstates are
built purely from spacetime fields and not the equations of motion used to
produce those fields. Yet, the domains and structures they capture are
remarkably close to the dynamical systems benchmark set by \DPID. This is
highly encouraging as the \localstates can be uniformly applied to a much wider
array of systems than the \DPID domain transducers, while at the same
time providing a more powerful analysis of coherent structures.

Looking beyond cellular automata, recent years witnessed renewed interest in
coherent structures in fluid systems \cite{Hall15a, Mezi13a, Holm12a}. There
has been particular emphasis on Lagrangian methods, which focus on material
deformations generated by the flow. The \localstates, in contrast, are an
Eulerian approach, as they are built from lightcones taken from spacetime
fields and do not require material transport in the system. A frequent
objection raised against Eulerian approaches to coherent structures is that
such approaches are not ``objective''---they are not independent of an
observer's frame of reference. This applies for instantaneous Eulerian
approaches, however. And so, does not apply to \localstates. In fact,
lightcones and the local causal equivalence relation over them are preserved
under Euclidean isometries. This can be seen from Eqs. (\ref{eq:plc}) and
(\ref{eq:flc}) that define lightcones in terms of distances only and so they
are independent of coordinate reference frame. \Localstates are objective in
this sense.

Methods in the Lagrangian coherent structure literature fall into two main
categories: diagnostic scalar fields and analytic approaches utilizing one or
another mathematical coherence principle. Previous approaches to coherent
structures using \localstates relied on the local statistical complexity
\cite{Shal06a,Jani07a}. This is a diagnostic scalar field and comes with all
the associated drawbacks of such approaches \cite{Hadj17a}. The coherent
structure theory presented here, in contrast, is the first principled
mathematical approach to coherent structures using \localstates.

With science producing large-scale, high-dimensional data sets at an ever
increasing rate, data-driven analysis techniques like the \localstates become
essential. Standard machine learning techniques, most notably deep learning
methods, convolutional neural nets, and the like are experiencing increasing
use in the sciences \cite{Liu16a, Bald014a}. Unlike commercial applications in
which deep learning has led to surprising successes, scientific data is highly
complex and typically unlabeled. Moreover, interpretability and detecting new
mechanisms are key to scientific discovery. With these challenges in mind, we
offer \localstates as a unique and valuable tool for discovering and
understanding emergent structure and pattern in spatiotemporal systems.

\section*{Acknowledgments}
\label{sec:acknowledgments}

The authors thank Bill Collins, Ryan James, Karthik Kashinath, John Mahoney, Mr.
Prabhat, Paul Riechers, Anastasiya Salova, and Dmitry Shemetov for helpful discussions and feedback,
and the Santa Fe Institute for hospitality during visits. JPC is an SFI External
Faculty member. We thank Ryan James and Dmitry Shemetov for help with software
development. This material is based upon work supported by, or in part by, the
John Templeton Foundation grant 52095, Foundational Questions Institute grant
FQXi-RFP-1609, and the U. S. Army Research Laboratory and the U. S. Army Research
Office under contract W911NF-13-1-0390, as well as by Intel through its support
of UC Davis' Intel Parallel Computing Center.

%
%
%
%
%
%
%
%
%
%
%
%
%
%
%
%
%
%
%
%
%
%
%
%
%
%
%
%
%
%
%
%
%
%
%
%
%
%
%
%
%
%
%
%
%
%
%
%
%
%
%
%
%
%
%
%
%
%
%
%
%
%
%
%
%
%
%
%
%
%
%
%
%
%
%
%
%
%
%
%
%
%
%
%
%
%
%
%
%
%
%
%
%
%
%
%
%
%
%
%
%
%
%
%
%
%
%
%
%
%
%
%
%
%
%
%
%
%
%
%
%
%
%
%
%
%
%
%
%
%
%
%
%
%
%
%
%
%
%
%
%
%
%
%
%
%
%
%
%
%
%
%
%
%
%
%
%
%
%
%
%
%
%
%
%
%
%
%
%
%
%
%
%
%
%
%
%
%
%
%
%
%
%
%
%
%
%
%
%
%
%
%
%
%
%
%
%
%
%
%
%
%
%
%
%
%
%
%
%
%
%
%
%
%
%
%
%
%
%
%
%
%
%
%
%
%
%
%
%
%
%
%
%
%
%
%
%
%
%
%
%
%
%
%
%
%
%
%
%
%
%
%
%
%
%
%
%
%
%
%
%
%
%
%
%
%
%
%
%
%
%
%
%
%
%
%
%
%
%
%
%
%
%
%
%
%
%
%
%
%
%
%
%
%
%
%
%
%
%
%
%
%
%
%
%
%
%
%
%
%
%
%
%
%
%
%
%
%
%
%
%
%
%
%
%
%
%
%
%
%
%
%
%
%
%
%
%
%
%
%
%
%
%
%
%
%
%
%
%
%
%
%
%
%
%
%
%
%
%
%
%
%
%
%
%
%
%
%
%
%
%
%
%
%
%
%
%
%
%
%
%
%
%
%
%
%
%
%
%
%
%
%
%
%
%
%
%
%
%
%
%
%
%
%
%
%
%
%
%
%
%
%
%
%
%
%
%
%
%
%
%
%
%
%
%
%
%
%
%
%
%
%
%
%
%
%
%
%
%
%
%
%
%
%
%
%
%
%
%
%
%
%
%
%
%
%
%
%



\end{document}